\documentclass[aps,pra,twocolumn,superscriptaddress]{revtex4-1}
\usepackage{graphicx}% Include figure files
\usepackage{dcolumn}% Align table columns on decimal point
\usepackage{bm}% bold math
\usepackage{color}
\usepackage{multirow}
\usepackage[mathscr]{euscript}
\usepackage{amsmath}

\usepackage[T1]{fontenc}
\usepackage{tikz}
\newcommand*\circled[1]{\raisebox{.9pt}{\textcircled{\raisebox{-.8pt}{#1}}}}

\usepackage[colorlinks=true,allcolors=blue]{hyperref}

\begin{document}
\title{Quench dynamics of Rydberg-dressed bosons on two-dimensional square  lattices}

\author{Yijia Zhou}
\affiliation{School of Physics and Astronomy, University of Nottingham, Nottingham NG7 2RD, United Kingdom}
\author{Yongqiang Li}
\affiliation{Department of Physics, National University of Defense Technology, Changsha 410073, P. R. China}
\affiliation{Department of Physics, Graduate School of China Academy of Engineering Physics, Beijing 100193, P. R. China}
\author{Rejish Nath}
\affiliation{Indian Institute of Science Education and Research, Pune 411 008, India}
\author{Weibin Li}
\affiliation{School of Physics and Astronomy, University of Nottingham, Nottingham NG7 2RD, United Kingdom}
\affiliation{Centre for the Mathematics and Theoretical Physics of Quantum Non-Equilibrium Systems, University of Nottingham, Nottingham NG7 2RD, United Kingdom}

\begin{abstract}
We study the dynamics of bosonic atoms on a two-dimensional square lattice, where atomic interactions are long-ranged with either a box or soft-core shape. The latter can be realized through laser dressing ground-state atoms to electronically excited Rydberg states. When the range of interactions is equal or larger than the lattice constant, the system is governed by an extended Bose-Hubbard model. We propose a quench process by varying the atomic hopping linearly across phase boundaries of the Mott insulator-supersolid and supersolid-superfluid phases. Starting from a Mott insulating state, the dynamical evolution of the superfluid order parameter exhibits a universal behaviour at the early stage, largely independent of interactions. The dynamical evolution is significantly altered by strong, long-range interactions at later times. Particularly, we demonstrate that density wave excitation is important when the quench rate is small. Moreover, we show that the quench dynamics can be analyzed through time-of-flight images, i.e., measuring the momentum distribution and noise correlations. 
\end{abstract}
\date{\today}
\keywords{}
\maketitle

\section{Introduction}

In the past decades, there has been a growing interest in the study of ultracold atoms, which is largely driven by the unprecedented level of control over external trapping potentials, internal states and interactions between atoms with electromagnetic fields~\cite{Jaksch1998, Stenger1999, Theis2004, Chin2010, Schachenmayer2010, Johnson2010, Anderson2011, Viteau2011, Macri2014}. Various lattice models~\cite{Gross2017}, such as the Bose-Hubbard model~\cite{Fisher1989}, have been studied and realized experimentally~\cite{Greiner2002a}. This has opened opportunities to probe static properties~\cite{Jaksch2005, Lewenstein2007, Bloch2008}, such as Mott insulator-superfluid phase transition~\cite{Fisher1989, VanOosten2001, Greiner2002a, Zwerger2003}, spin-orbit coupling \cite{Wall2016, Bromley2016, Zhang2018}, supersolidity \cite{Sengupta2005, Scarola2006, Batrouni2006, Menotti2007, Yi2007, Iskin2009, Iskin2011, Li2012, Landig2016, Li2018}, entanglement \cite{Alba2013, Lukin2015}, topology \cite{Goldman2016, Lohse2016}, etc. In cold atom systems, many parameters can be manipulated and monitored dynamically. This allows exploring non-equilibrium dynamics in addition to steady states. Theoretical and experimental works have investigated Landau-Zener transitions~\cite{Wu2003, Tomadin2008, Deng2015a}, Kibble-Zurek mechanism \cite{Zurek2005, Dziarmaga2012, DelCampo2013, Shimizu2018a, Shimizu2018b, Shimizu2018c, Weiss2018}, transport  \cite{Scherg2018, Brown2018, Fujiwara2019}, and excitations of Higgs and Goldstone modes \cite{Leonard2017a, DiLiberto2018}.

\begin{figure}
	\centering
	\includegraphics[width=0.9\linewidth]{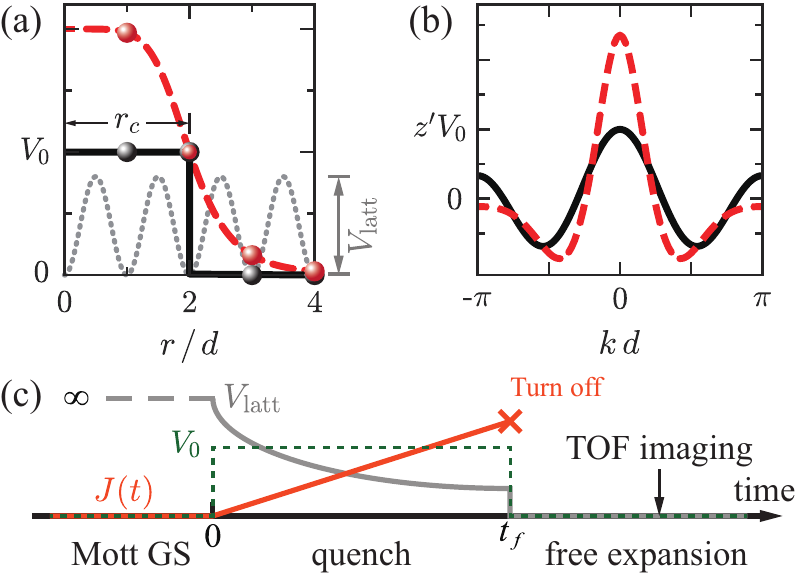} 
	\caption{\textbf{Long-range interaction and quench scheme.} (a) Long-range soft-core (red dashed) and box (black solid) interaction. The range $r_c$ of interactions can be larger than the optical lattice (dotted) constant $d$. (b) Fourier transform of the interaction potential. The line style is the same as shown in panel (a). (c) Quench protocol. At $t<0$, we prepare the ground state in a Mott insulator with $V_{\rm latt}\neq 0$ and $J=0$. The tunnelling $J(t)/U=a_Q t$ (orange) is increased linearly during $t\in(0,t_f)$. This is done by reducing lattice depth (grey). At the same time, long-range interactions (green dotted) are turned on. When $t>t_f$, atoms are released from the optical lattice to initialize the time-of-flight experiment.}
\label{fig:intro}
\end{figure}

Recently, growing interest has been spent on investigating the dynamics of Bose-Hubbard models (BHMs) driven by external periodic ~\cite{Eckardt2005, Gaul2009} or linear fields~\cite{Green2005}. A recent review can be found in Refs.~\cite{Polkovnikov2011, Kennett2013}. A particularly interesting topic is the universal dynamics found in BHMs when linearly changing the hopping strength.  Due to the Kibble-Zurek mechanism (KZM)~\cite{Zurek2005, Dziarmaga2012, DelCampo2013},  the dynamics is frozen around the phase boundary while adiabatic away from it. Many quantities, such as correlation lengths and topological defects, exhibit universal behaviours.
 
In BHMs, the competition between the hopping and two-body on-site interactions~\cite{Freericks1994, Elesin1994, Folling2005}  leads to a Mott insulator (MI) to superfluid (SF) phase transition at some critical $J/U$. When the interaction length is greater than the lattice constant $d$, we obtain an extended Bose-Hubbard model (eBHM). Its ground state can have non-uniform, periodic densities, such as the density wave (DW), supersolid (SS), Haldane insulator, etc.~\cite{Kuhner2000, Rossini2012}. Such a situation has been examined extensively using atoms with weak magnetic or electric dipole moments~\cite{Goral2000, Griesmaier2005, Lu2011, Aikawa2012, Buchler2007, Micheli2006, Kotochigova2006, Gorshkov2011, Baier2016}, where the dominant interaction is between the nearest-neighbour sites. Drastically different dynamics is found when quenching the eBHMs~\cite{Shimizu2018b, Shimizu2018c}. 

In this work, we will go beyond the nearest-neighbour interaction regime and examine the dynamics of eBHMs with even longer-range interactions. This situation can be realized by using Rydberg-dressed atoms confined in a two-dimensional optical lattice~\cite{Balewski2014, Macri2014, Jau2016, Zeiher2016, Mukherjee2016, Koscik2018, Koscik2019}. Rydberg atoms have long lifetimes $10\sim 100~\mu\text{s}$, growing with  $n^3$ where the principal quantum number $n\gg 1$. Large polarizabilities ($\propto n^{7}$) give strong and long-range interactions (e.g. van der Waals interactions $\propto n^{11}$) between Rydberg atoms. However, atomic hopping typically occurs on a much slower time scale than the Rydberg lifetime. One could instead weakly couple ground-state atoms to Rydberg states with a far detuned laser~\cite{Henkel2010, Honer2010, Johnson2010, Pupillo2010, Chougale2016}. This results in a soft-core shape potential between Rydberg dressed ground-state atoms. Its range $r_c$ could extend several lattice constants before decaying significantly (see Fig.~\ref{fig:intro}a). The interaction strength is nearly a constant for distance $r<r_c$ and decays quickly when $r>r_c$. Such an interaction potential may be approximated by a box potential, i.e. atomic interaction is a constant when $r\le r_c$ and zero otherwise (Fig.~\ref{fig:intro}a). Box-type interactions have been used to study the extended Bose-Hubbard model, with a focus on nearest-neighbour interactions ($r_c=d$)~\cite{Kuhner2000, Iskin2011, Rossini2012}. As the two types of potential share similar profiles in momentum space (Fig.~\ref{fig:intro}b), we will show that the respective dynamics shows common features. 

This paper is organized as follows. In Sec. II we introduce the eBHM of the Rydberg dressed atoms. Ground-state phase diagrams are calculated using the Gutzwiller method. We show the relation between the roton instability and density modulation by analysing the Bogoliubov spectra. In Sec. III, we discuss universal dynamics for the box interaction in the MI-SF phase transition and dynamics in the SS and SF phases. In Sec. IV, we propose that the quench dynamics can be measured through time-of-flight (TOF) density distributions~\cite{Greiner2002b, Li2012, Baier2016} and covariance~\cite{Altman2004, Folling2005, Rom2006, Jeltes2007, Toth2008, Hu2010} at different probing times. In Sec. V, dynamics for the soft-core interaction is discussed. It is found that the dynamics of the eBHM with the soft-core interaction is largely similar to the box interaction.  We conclude in Sec. VI. 

\section{The extended Bose-Hubbard Model and ground-state phase}

The Hamiltonian of the Rydberg dressed atoms in the 2D square lattice is given by an eBHM,
\begin{equation} \label{eq:eBHM}
H = - J(t) \sum_{\langle ij \rangle} \hat{b}_i^\dag \hat{b}_j + \frac{U}{2} \sum_i \hat{n}_i(\hat{n}_i-1) + \sum_{i \neq j} V_{ij} \hat{n}_i\hat{n}_j,
\end{equation}
where $\langle ij \rangle$ stands for the nearest-neighbour sites, and $J(t)$ is the time-dependent hopping. $U$ is the on-site interaction, including contributions from s-wave scattering and the long-range interaction.  In this work, two types of long-range interactions are considered. The soft-core interaction has a form $V_{ij} = 2V_0 [1+(r_{ij}/r_c)^6]^{-1}$, where $r_{ij}$ is the distance between site $i$ and $j$, $r_c$ is the soft-core radius, and $V_0$ is the interaction strength. In the case of a box potential, the interaction is given by $V_{ij} = V_0 \Theta(r_c - r_{ij})$, where the Heaviside function $\Theta(r)$ defines the box length $r_c$. Here we will use $r_c$ to denote the interaction range for both box and soft-core interactions.

We employ the Gutzwiller approach to calculate the ground state and dynamics of the Hamiltonian~\cite{Krauth1992, Seibold2001, Schiro2010, VonOelsen2011a}. The Gutzwiller approach is a mean-field method and predicts qualitatively accurate phase boundaries. Decoupling the many-body wave function into local wave functions $|\Psi_N\rangle \approx \Pi_i|\Psi_i\rangle$, where the  $i$-th site wave function is expanded using Fock state basis $|m\rangle$ ($m=0,1,\cdots$) as $|\Psi_i\rangle = \sum_m f_m^i(t) |i;m\rangle$ with $|i;m\rangle$ to represent $m$ atoms in $i$-th site and $f_m^i(t)$ to be the corresponding probability amplitude. The equation of motion of coefficients $f_m^i$ is 
\begin{eqnarray} \label{eq:EOM} 
i\frac{\partial}{\partial t} f_m^i &=&
-J(t) \sum_{r_{ij}=1} \Bigg( \sqrt{m} \phi_j f_{m-1}^i + \sqrt{m+1} \phi_j^* f_{m+1}^i \Bigg) \nonumber \\
&\phantom{={}}& + \Bigg( \frac{U}{2} m(m-1) + m \sum_{j\neq i} V_{ij} n_j \Bigg) f_m^i,
\end{eqnarray}
where $\phi_i=\langle \hat{b}_i \rangle$ is the SF order parameter. 

We find the ground state by solving Eq.~(\ref{eq:EOM}) in the imaginary time via the Nelder-Mead algorithm. This is done with $32\times32$ sites and the maximal occupation in each site is 12. In the MI phase, particle numbers at each site are integer and SF order parameter $\phi_i=0$. The SF phase is determined by a non-zero, uniform SF order parameter $\phi_i=\phi$. The static structure factor is used to identify density-modulated phases, 
\begin{equation}
S(\mathbf{k})=\frac{1}{M^2}\sum_{j,l}e^{i\mathbf{k}\cdot(\mathbf{r}_j-\mathbf{r}_l)}\langle \hat{n}_j\hat{n}_l\rangle,
\end{equation}
where $\mathbf{k}\equiv (k_x,k_y)$ is the reciprocal lattice vector (in terms of $1/d$), and $M$ is the total number of lattices~\cite{bandyopadhyay_quantum_2019}. In DW states the structure factor $S(\pi,\pi)\neq 0$ and SF order parameter $\phi_i=0$. In SS phases, both SF order parameter and the structure factor are non-zero. 

\begin{figure}
	\centering
	\includegraphics[width=\linewidth]{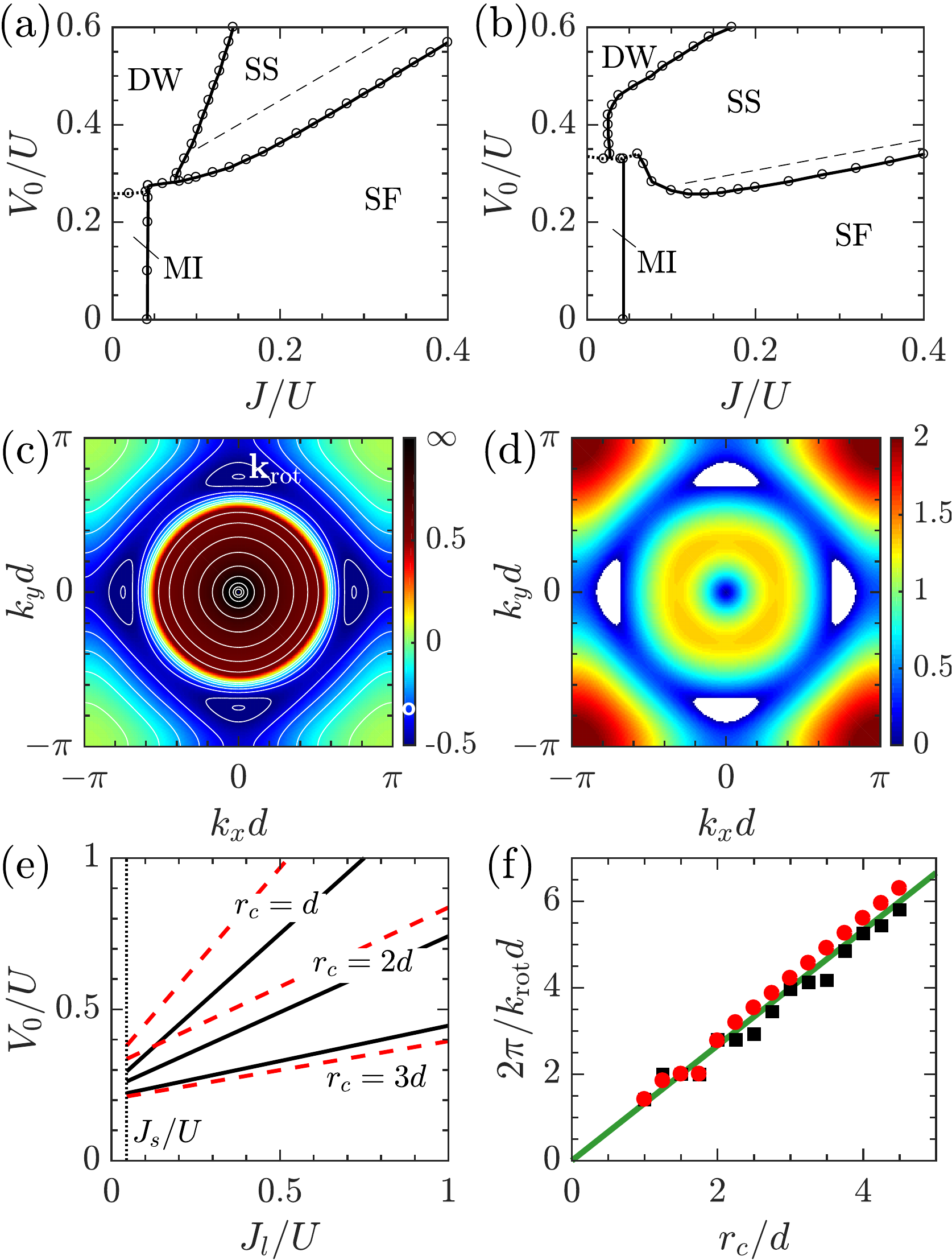} 
	\caption{\textbf{Phase diagram and roton instability.} Phase diagram for the box-type interaction with $r_c=d$ (a) and for the soft-core interaction with $r_c=2.5~d$ (b) at unit filling. When the long-range interaction is weak, a MI-SF transition is found by increasing $J$. Starting from the DW phase (at stronger $V_0$), the ground state undergoes a DW-SS and then a SS-SF transition when increasing $J$. The dashed line is calculated from the roton instability analysis, which is close to the SS-SF phase boundary.  (c) $\Sigma(\mathbf{k})$ as defined in Eq.~\eqref{eq:Jcl}. The momentum $\mathbf{k}_{\rm rot}$ is found at the minimal $\Sigma(\mathbf{k})$ when the spectra become complex. (d) Bogoliubov spectra for $J=0.5U$. The white areas indicate roton instability. In (c) and (d) we consider the soft-core interaction with $r_c=2d$ and $V_0=0.6U$. (e) Critical tunnelling $J_l$ for box-type (black solid) and soft-core (red dashed) interactions. The interaction lengths are $r_c=\{d,2d,3d\}$, respectively. (f) Momentum $|\mathbf{k}_{\text{rot}}|$ at the roton minimum, in term of its inverse, for box (square) and soft-core (dot) interactions. Here the interaction strength is $V_0=2U$. }
	\label{fig:bogoliubov}
\end{figure}

Ground-state phase diagrams for the box interaction have been examined by similar methods (see, e.g. Ref.~\cite{Iskin2009}). It was shown that the ground state exhibits MI, DW, SS and SF phases by varying hopping, on-site and long-range interactions. We calculate the phase diagram with average particle number in each site $n_i=1$. One example for the box interaction with $r_c=d$ is shown in Fig.~\ref{fig:bogoliubov}a. When the tunnelling and $V_0$ are weak, the ground state is a MI. Starting from the MI, the system undergoes a MI-SF phase transition when $J$ increases. With larger $V_0$, the ground state enters DW phases when the hopping is small.  We find a DW-SS and then SS-SF transition by increasing hopping $J$~\cite{bandyopadhyay_quantum_2019}. More cases and discussions of the phase diagram can be found in ~\cite{Iskin2009}. In the case of soft-core interactions, the phase diagram has similar distributions of phases. In Fig.~\ref{fig:bogoliubov}a, we give one example for $r_c=2.5d$. In this case, the SS phase region becomes larger in the parameter space due to the longer-range interaction.

The emergence of the SS phase can be further analyzed using the Bogoliubov theory. In the SF region, the ground state is a homogeneous condensate, while SS leads to spatial modulations in the condensate. When this happens, the Bogoliubov spectrum of the homogeneous SF phase~\cite{Rey2003, Macri2013}, 
\begin{equation}
E(\mathbf{k}) = \sqrt{ \varepsilon(\mathbf{k})^2 + \rho_0 \varepsilon(\mathbf{k}) \left( U+\tilde{V}(\mathbf{k}) \right) },
\end{equation}
becomes unstable. Here $ \varepsilon(\mathbf{k}) = 2J \big[2 - \cos(k_x d) - \cos(k_y d) \big]$ is the kinetic energy term, $\rho_0$ is the condensate density and $\tilde{V}(\mathbf{k}) = \frac{1}{M}\sum_{i \neq j} V_{ij} \exp(i\mathbf{k}\cdot\mathbf{r}_{ij})$, is the Fourier transformation of the long-range interaction. In free space, $\tilde{V}(\mathbf{k})$ is calculated analytically, which yields $\tilde{V}(\mathbf{k})=V_0 \frac{2\pi r_c}{|\mathbf{k}|} J_1(|\mathbf{k}| r_c)$ for the box interaction and $\tilde{V}(\mathbf{k})=V_0\frac{4\pi r_c^2}{3} \sum_{m=0,\pm1} e^{i 2m\pi/3} K_0(e^{i m\pi/3}|\mathbf{k}| r_c) $ for the soft-core interaction, where $J_n(x)$ is the first kind of Bessel function and $K_n(x)$ is the second kind of modified Bessel function \cite{Hsueh2012}. In the two-dimensional lattice, the Fourier transform of the interaction potential is numerically calculated and one example is presented in Fig.~\ref{fig:intro}b. It is found that $\tilde{V}(\mathbf{k})$ is negative for a large range of $kd$. 

When the long-range interaction is strong,  the negative components of $\tilde{V}(\mathbf{k})$ cause a roton instability. Here the Bogoliubov spectra $E(\mathbf{k})$ becomes imaginary. The critical hopping (i.e. the lower bound) to observe the roton instability is $J_l=-\min[\Sigma(\mathbf{k})]$, where
\begin{equation} \label{eq:Jcl}
\Sigma(\mathbf{k}) = \rho_0 \frac{U+\tilde{V}(\mathbf{k})}{ 2-\cos(k_x d)-\cos(k_y d) }.
\end{equation}
A typical $\Sigma(\mathbf{k})$ for the soft-core interaction with $r_c=d$ and $V_0=0.6U$ is shown in Fig.~\ref{fig:bogoliubov}c. When $J=0.5U$, the roton instabilities are triggered and found in the white area in the Bogoliubov spectrum (Fig.~\ref{fig:bogoliubov}d). 
Through numerically fitting $J_l$ versus $V_0$, its slope for the soft-core interaction becomes larger than that of the box interaction when increasing $r_c$ (Fig.~\ref{fig:bogoliubov}e). On the other hand, momentum $\mathbf{k}_{\rm rot}$ corresponding to the onset of roton instabilities varies with $r_c$. From the numerical data, one finds that $|\mathbf{k}_{\rm rot}| \approx 3\pi/2r_c$ when $r_c\gg d$ (Fig.~\ref{fig:bogoliubov}f).  

\section{Quench Dynamical with box interaction}
When $t<0$, the long-range interaction is not present yet. We prepare the system in the MI state with mean particle number $ n_i =\langle b_i^{\dagger}b_i\rangle=1$. When $t>0$, the box interaction is turned on instantaneously and the hopping is increased linearly, i.e., $J(t)=a_{Q} t$ with $a_{Q}$ being the quench rate. The time sequence is shown in Fig.\ref{fig:intro}c. 
Response of the system is captured by average quantities such as superfluid fraction $\rho_s =  \sum_{\langle i,j \rangle} {\rm Re}[\phi_i^*\phi_j]/zM $~\cite{Roth2003, Damski2003}, density variance $\sigma_n = \sum_i \sqrt{\langle{\hat{n}_i}^2\rangle - \langle{\hat{n}_i}\rangle^2}/M$ and vortex nucleation. Here $z=4$ is the coordination number of the 2D square lattice and $M$ is the total number of lattice points. In the MI phase ($J(t)/U\ll 1$), both $\rho_s$ and $\sigma_n$ vanish. 
\begin{figure}
	\centering
	\includegraphics[width=\linewidth]{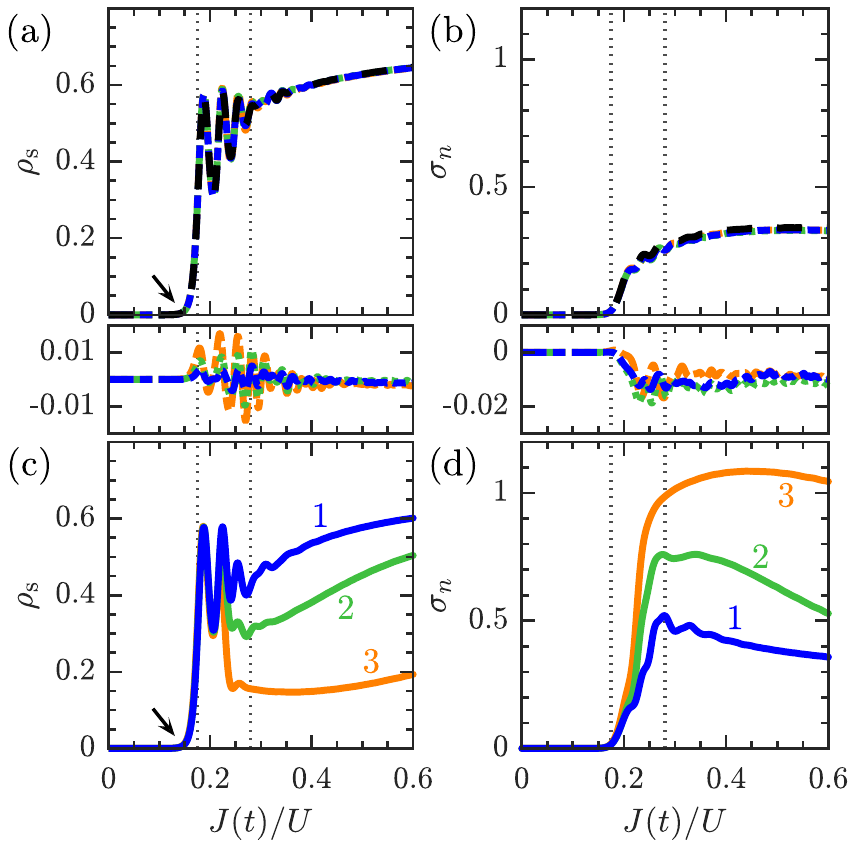} 
	\caption{\textbf{Superfluid fraction $\rho_s$ (a,c) and density variance $\sigma_n$ (b,d).}  In weak interacting cases (a,b), we consider $V_0=0$ (dashed black) and $V_0/U=0.1$ (colored). The lower panel shows the difference of $\rho_s$ and $\sigma_n$ between interacting and non-interacting cases. In strong interacting cases with $V_0/U=0.5$ (c,d), the evolution apparently depends on the strength and radius of the long-range interaction. In the simulation, $a_Q/U=0.01$ and $r_c/d=\{1$ (blue), $2$ (green), $3$ (orange)$\}$. Dotted lines indicate probe times for time-of-flight (TOF) interference. Arrows indicate the Kibble-Zurek time. See text for details.}
	\label{fig:Box_rho_dn}
\end{figure}

The dynamics of the eBHM is obtained by numerically solving Eq.~(\ref{eq:EOM}) using the fourth-order-Runge-Kutta algorithm. In the initial state $f_1^i \approx e^{i\theta_i}$ where $\theta_i$ is a random phase that uniformly distributes from $0$ to $2\pi$. We then enforce particle number fluctuations in the order of $10^{-3}$ to $m\neq 0$ states and the normalization condition $\sum_m|f_m^{j}|^2=1$. The lattice size is chosen to be $128\times128$ for box interactions and $48\times48$ for soft-core interactions with a maximal occupation 7 for each site. The dynamics remains largely the same if we increase the number of sites. We cast 40 trajectories for different initial states and evaluate physical quantities by calculating their average values. 

\subsection{Superfluid fraction, density variance and vortex density}
When the box interaction is weak, the dynamics is largely similar to that of the BHM, as shown in Fig.~\ref{fig:Box_rho_dn}a,b. Initially both $\rho_s$ and $\sigma_n$ remain small with increasing $J(t)$. During this stage, atomic tunnelling is negligible until $J(t)$ increases to $\sim 0.17U$. This forms the early stage of evolution. At the second stage $J(t)>0.17U$,  $\rho_s$ and $\sigma_n$ increase rapidly. Here the long-range interaction acts as a weak perturbation on top of the strong on-site interaction (see lower panel of Fig.~\ref{fig:Box_rho_dn}a,b, where the difference between interacting and noninteracting case is plotted). The value $J/U\approx 0.17$ is related to the Kibble-Zurek mechanism, which will be discussed in the next section. 
\begin{figure}
	\centering
	\includegraphics[width=0.95\linewidth]{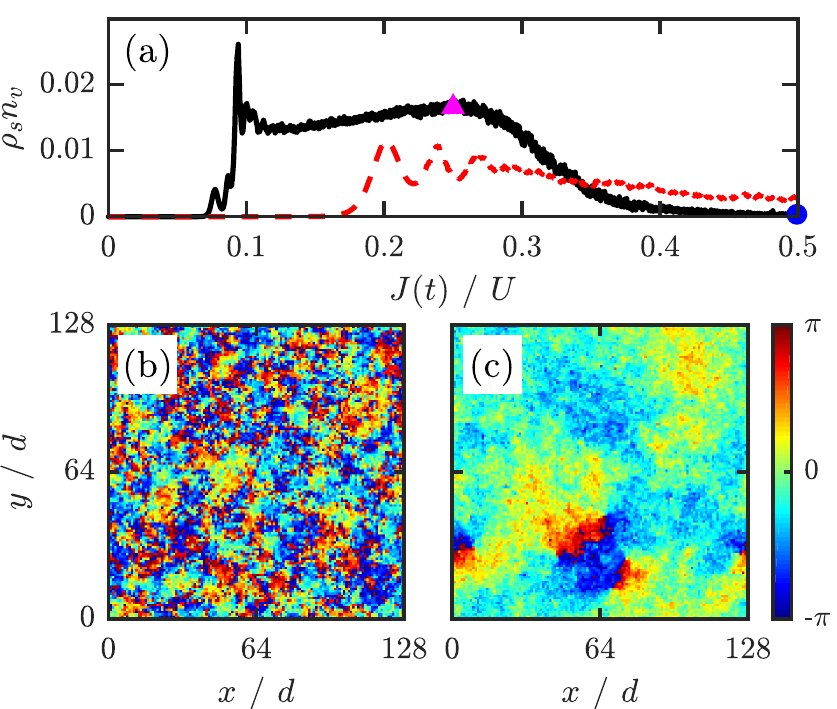} 
	\caption{\textbf{Dynamics of the vortex nucleation.} (a) Evolution of superfluid vortex density for box interactions with $V_0/U=0.5$, $r_c=d$ and $J(t_f)/U=0.5$. The quench rate $a_Q/U$ is $10^{-3}{\rm\ (black),\ }10^{-2}{\rm\ (red)}$. The local phase of SF order parameter for quench labeled as magenta triangle and blue circle is shown for (b) $J(t)/U=0.25$ and (c) $J(t)/U=0.5$.}
	\label{fig:vortices}
\end{figure}

\begin{figure*}
\centering
\includegraphics[width=0.95\linewidth]{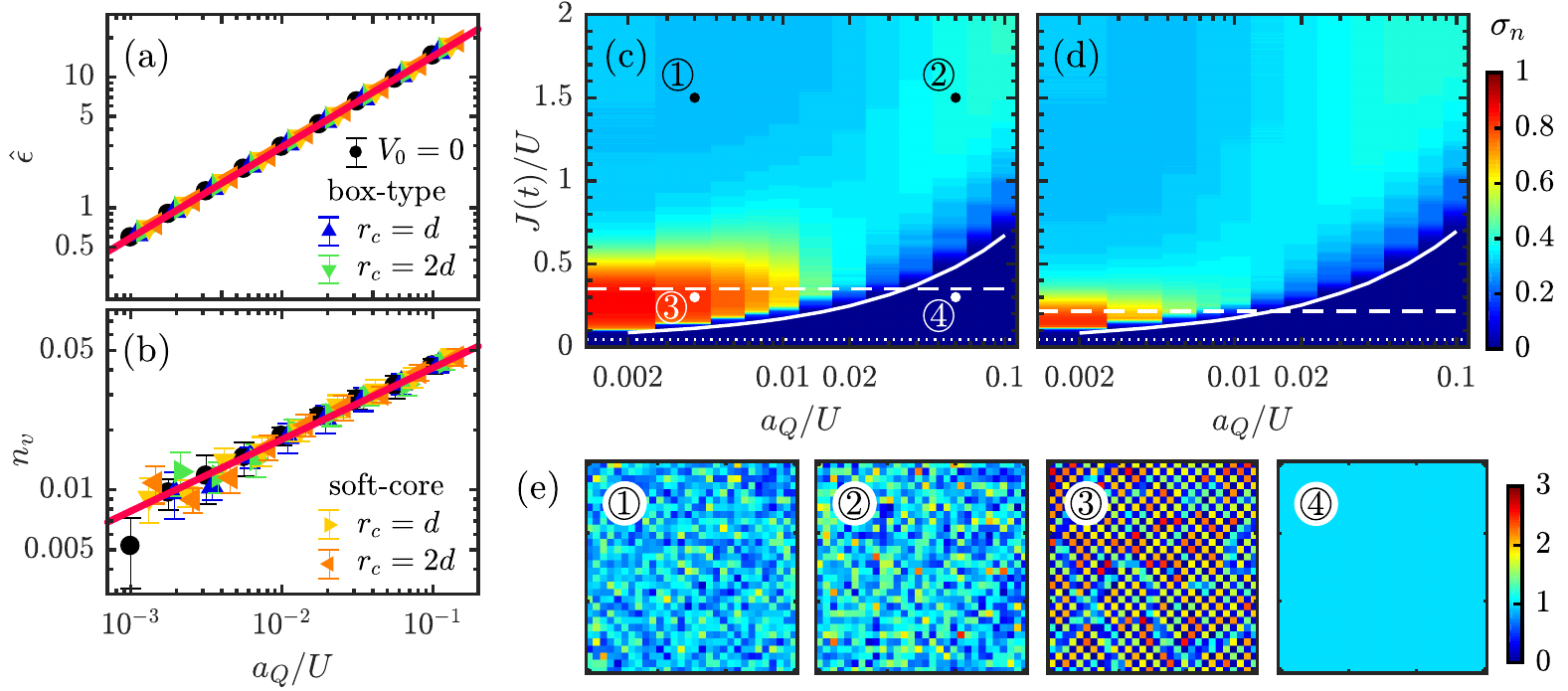} 
\caption{\textbf{Kibble-Zurek time scale.} Dynamics of global variables for box and soft-core interactions.(a) Kibble-Zurek time, in term of frozen parameter $\hat{\epsilon}$, fitted from the turning point of superfluid fraction. The interaction configurations include non-interacting (black), box interaction with $V_0/U=0.6$ and $r_c=d$ (blue), $r_c=2d$ (green), soft-core interaction with $V_0/U=0.6$ and $r_c=d$ (yellow), $r_c=2d$ (orange). (b) Density of vortices counted at Kibble-Zurek time. Density variance $\sigma_n$ with respect to hopping $J(t)$ with box (c) and soft-core (d) interaction for both $r_c=d$ and $V_0/U=0.6$. The hopping parameter at KZM time $J(\hat{t})$ (solid line), roton instability $J_l$ (dashed line) and MI-SF transition of BHM $J_s$ (dotted line) are plotted together. (e) Density distributions of specific quench rate and time for box interactions shown in (c), where \circled{1} $a_Q/U=0.004$, $J(t)/U=1.5$, \circled{2} $a_Q/U=0.06$, $J(t)/U=1.5$, \circled{3} $a_Q/U=0.004$, $J(t)/U=0.3$ and \circled{4} $a_Q/U=0.06$, $J(t)/U=0.3$. }
\label{fig:KZM_fit}
\end{figure*}

Dynamics changes drastically at the two stages when the long-range interaction becomes stronger, as illustrated in Fig.~\ref{fig:Box_rho_dn}c,d. At early times, SF density $\rho_s$ is independent of the long-range interaction. However, $\rho_s$ decrease with increasing $r_c$ at later times. The density variance $\sigma_n$ is more sensitive to $r_c$. Different curves depart from each other when $J(t)>0.17U$. From a mean-field level, we can understand these curves by noting that the density-density interaction at a given site becomes stronger due to the long-range interaction (i.e., with multiple sites). Eventually the local SF order parameter is reduced with larger soft-core radius (Fig.~\ref{fig:Box_rho_dn}c). At the same time, the density fluctuations are enhanced (Fig.~\ref{fig:Box_rho_dn}d), as density waves are excited during the quench (see examples in Fig.~\ref{fig:KZM_fit}).

When $J(t)$ is increased, topological defects (vortices) can be created in the many-body state. Numerically, number densities of vortices $n_v$ are evaluated according to $n_v=1/(2\pi M)\sum_i |\arg(\phi_{i+\hat{x}}^* \phi_{i}) + \arg(\phi_{i+\hat{x}+\hat{y}}^*\phi_{i+\hat{x}}) + \arg(\phi_{i+\hat{y}}^*\phi_{i+\hat{x}+\hat{y}}) +\arg(\phi_{i}^*\phi_{i+\hat{y}}) |$ with $\hat{x}$ and $\hat{y}$ to be the unit vector along $x$ and $y$ axis. The development of the topological defects in the vicinity of the phase transition point is related to quantities such as the healing length, density fluctuation, etc., as found in BHMs~\cite{Dziarmaga2012} and eBHMs with dipolar interactions~\cite{Yi2007, Baier2016}. Due to the random phases in the initial state, there are vortices even before the SF order parameter develops. We thus will have phase winding but no currents. To exclude this situation, we define a superfluid vortex density $\rho_s n_v$, which takes into account contributions from both the SF density and vortex density.

For intermediate quench rate $a_Q/U=0.01$, the superfluid vortex density increases rapidly around $J(t)/U=0.17$, as shown in  Fig.~\ref{fig:vortices}a. In this case, there will be some vortices even when $J(t)$ is in the SF phase region (Fig.~\ref{fig:vortices}b). When the quench rate is further decreased to $a_Q/U= 10^{-3}$, the vortex density increases quickly when $J(t) > 0.071U$. This comes from the fact that low-energy modes are excited during the slow quench, which create many vortices. Subsequently, the SF vortex density decreases with increasing $J(t)$. When $J(t)$ is large, vortex and anti-vortex pairs recombine at a higher rate, which speeds up the relaxation of the system to a homogeneous SF with nearly homogeneous phases (Fig.~\ref{fig:vortices}c). 

\subsection{Kibble-Zurek dynamics}
The key feature of the KZM in a BHM is that dynamics is divided into frozen and adiabatic region across the phase transition. It is convenient to define a \textit{distance parameter} $\epsilon = J(t)/J_s-1$, which depends on the critical hopping $J_s$ of the MI-SF transition. Dynamics is qualitatively different before and after a frozen parameter $\hat\epsilon$. As the instantaneous relaxation time $\tau(t)$ diverges around the phase transition, the dynamics is frozen to the initial state if $|\epsilon|<\hat\epsilon$. However, dynamical evolution becomes adiabatic when $|\epsilon|>\hat{\epsilon}$, where many dynamical quantities change significantly. The relaxation time can be estimated as $\tau=|\epsilon/\dot{\epsilon}|$. On the other hand, Landau's mean-field theory predicts that $\tau=\tau_0|\epsilon|^{-z\nu}$, where $\nu$ is the critical exponent about correlation length, $\tau_0$ is a coefficient of the relaxation time for a BHM, and $z$ is the dynamical exponent. The frozen parameter and frozen (KZM) time are~\cite{Dziarmaga2012}, 
\begin{equation} \label{eq:scaling}
\hat{\epsilon} = \left( \frac{a_Q\tau_0}{J_s/U} \right)^{\frac{1}{1+z\nu}}  \quad {,} \quad  \hat{t} = \frac{J_s/U}{a_Q} \left( \frac{a_Q\tau_0}{J_s/U} \right)^{\frac{1}{1+z\nu}}.
\end{equation}
In practice, it is difficult to determine the KZM time from numerical calculations. In this work, we estimate the time when ${\rm d}^2\rho_s/{\rm d}t^2$ is maximized (indicated by arrows in Fig.~\ref{fig:Box_rho_dn}a,c). In Shimizu, {\it et al}'s work the KZM time is chosen to be the time when $|\phi(t_0+\hat{t})|=2|\phi(t_0)|$ \cite{Shimizu2018a}. This makes minor differences in $\tau_0$, but the scaling exponents remain the same. 
\begin{figure}
	\centering
	\includegraphics[width=0.95\linewidth]{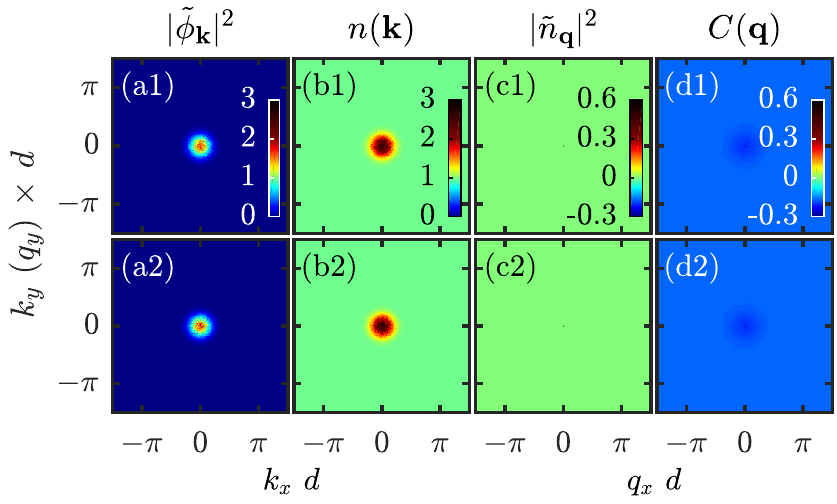} 
	\caption{\textbf{TOF snapshots at time $Ut=17.5$}. Before the atoms are released from the optical lattice, the interaction between them is a box type. The quench rate $a_Q/U=0.01$. TOF density $n(\mathbf{k})$ (b1-b2) and covariance $C(\mathbf{q})$ (d1-d2) are compared with the Fourier transform of order parameter $\tilde{\phi}_\mathbf{k}$ (a1-a2) and density $\tilde{n}_\mathbf{q}$ (c1-c2). Panels (a1-d1) show results of the BHM and  (a2-d2) results for box interaction with $r_c=d$ and $V_0/U=0.66$. }
	\label{fig:Compare_n_C_early}
\end{figure}

To find the universal behaviour in $\rho_s$, we change the quench rate from low to high, and find the respective KZM time. The corresponding frozen parameter $\hat{\epsilon}$ is shown in Fig.~\ref{fig:KZM_fit}a. It is found that $\hat{t}$  ($\hat{\epsilon}$) is largely independent to types of interactions. The data can be described by a single set of fitting parameters $J_s/U = 0.0449 $, $z\nu=0.442$ and $U\tau_0=21.0$. The resulting $J_s$ is close to the decoupling approach result $J_s/U=0.0429$ \cite{VanOosten2001} (Fig.~\ref{fig:bogoliubov}a,b). 

At the KZM time, the vortex density $n_v \propto \xi^{\sf def}/\xi^{\sf dim} \propto a_Q^{({\sf dim}-{\sf def})\nu/(1+z\nu)} $, where $\xi$ is the correlation length, and ${\sf dim}=2$ is the dimension of the lattice and ${\sf def}$ is that of topological defects. Here we argue that ${\sf def}=1$ because the vortices are always created in pairs (e.g., Fig.~\ref{fig:vortices}c).  From the numerical simulation, the scaling exponent $({\sf dim}-{\sf def})\nu/(1+z\nu)=0.359$ (see Fig.~\ref{fig:KZM_fit}b). Additionally, using the fitting results of $\hat{\epsilon}$, we obtain $\nu=0.518$ and $z=0.854$. Here one can make a comparison with mean-field results where $\nu=0.5$, $z=2$, and 3D $XY$ model, which is frequently compared with 2D BHM, where $\nu=2/3$ and $z=1$~\cite{Fisher1989, Elstner1999, Hohenadler2011}. In Ref.~\cite{Shimizu2018a}, it was found that the vortex density has anomalous behaviour in slow quench regime ($a_Q/U<10^{-2}$). In our simulations, $n_v$ shows larger fluctuation as $a_Q$ decreases due to the finite size effect.  The average values still agree with the universal scaling. 

The breakdown of the universal dynamics is more apparent for slow quenches, where the tunnelling $J(\hat{t})$ at the Kibble-Zurek time $\hat{t}$ is smaller than $J_l$. For fast quenches, however, the system will not fully respond to the roton mode, where dynamics is still universal. To demonstrate this, we calculate the density variance $\sigma_n$, shown in Fig.~\ref{fig:KZM_fit}c-e. The variance reaches a maximal value around $J(t)\sim J_l$ when the quench rate is low. In this region, the system suffers significantly from the roton instability, exhibiting apparent density patterns (Fig.~\ref{fig:KZM_fit}e3). 

\section{Time-of-flight analysis}
The phases shown above can be probed dynamically through analyzing the momentum distribution and noise correlation~\cite{Altman2004, Jeltes2007, Toth2008, Hu2010}. By releasing the optical lattice at different times, interference patterns shown in the time-of-flight images encode the phase information. In the deep SF regime, the momentum distribution $n(\mathbf{k})=\langle\hat{b}^\dag_\mathbf{k} \hat{b}_\mathbf{k}\rangle$, $\hat{b}_\mathbf{k}$ being the bosonic operator in momentum space, is approximately given by
\begin{equation} \label{eq:nk}
n(\mathbf{k}) \approx | \tilde{\phi}_{\mathbf{k}} |^2,
\end{equation}
where $\tilde{\phi}_{\mathbf{k}} = 1/\sqrt{M}\sum_i \phi_i \exp[i\mathbf{k}\cdot\mathbf{r}_i]$ is the Fourier component of the SF order parameter $\phi_i$.

Non-uniform density structures can be characterized e.g. by the noise correlation. This is done by calculating the covariance $C(\mathbf{k},\mathbf{k'}) = \langle \hat{b}^\dag_\mathbf{k} \hat{b}_\mathbf{k} \hat{b}^\dag_{\mathbf{k}'} \hat{b}_{\mathbf{k}'} \rangle - \langle \hat{b}^\dag_\mathbf{k} \hat{b}_\mathbf{k} \rangle \langle \hat{b}^\dag_{\mathbf{k}'} \hat{b}_{\mathbf{k}'} \rangle$. The covariance  can be obtained through HBT-type interference measurements. In the MI regime, the covariance is simplified to be ,
\begin{equation} \label{eq:Ck}
C(\mathbf{q}) \approx \left| \tilde{n}_{\mathbf{q}} \right|^2,
\end{equation}
where $\tilde{n}_{\mathbf{q}}$ is the Fourier component of occupation $n_i$ and the relative momentum $\mathbf{q}=\mathbf{k}'-\mathbf{k}$. The analytical expressions for $n(\mathbf{k})$ and $C(\mathbf{k},\mathbf{k}')$ for general situations are presented in the Appendix~A. Similar results can be found in Refs.~\cite{Altman2004, Folling2005, Toth2008}.
\begin{figure}
	\centering
	\includegraphics[width=0.95\linewidth]{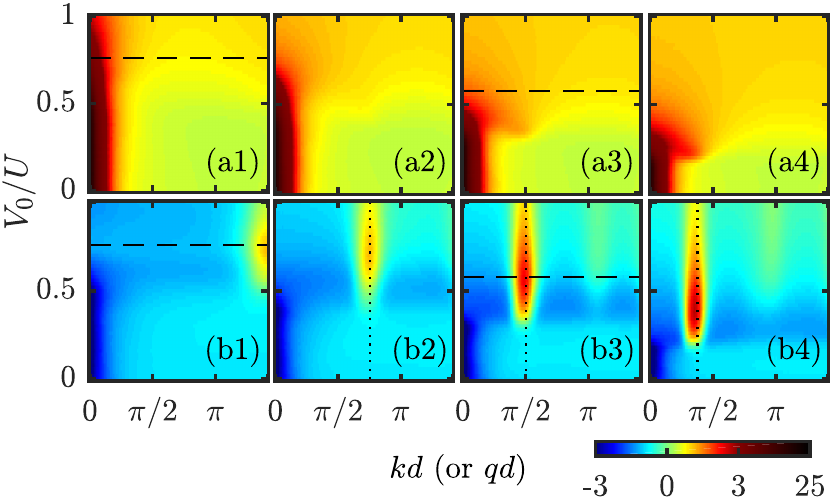} 
	\caption{\textbf{Radial distribution of the momentum density  and covariance. }  By integrating the angular direction, we obtain radial distribution of the momentum density $n(k)$ (a1-a4) and covariance $C(q)$ (b1-b4) at different interaction strengths. The probe time is $Ut=28$ with quench rate $a_Q/U=0.01$. We consider box interactions with radius $r_c/d=\{1,2,3,4\}$ from left to right, respectively. Dotted lines indicate the momentum where the covariance has maximal values.  See Fig.~\ref{fig:Compare_S_TOF} for snapshots of momentum distributions.   }
	\label{fig:Compare_n_C}
\end{figure}

\begin{figure}
	\centering
	\includegraphics[width=\linewidth]{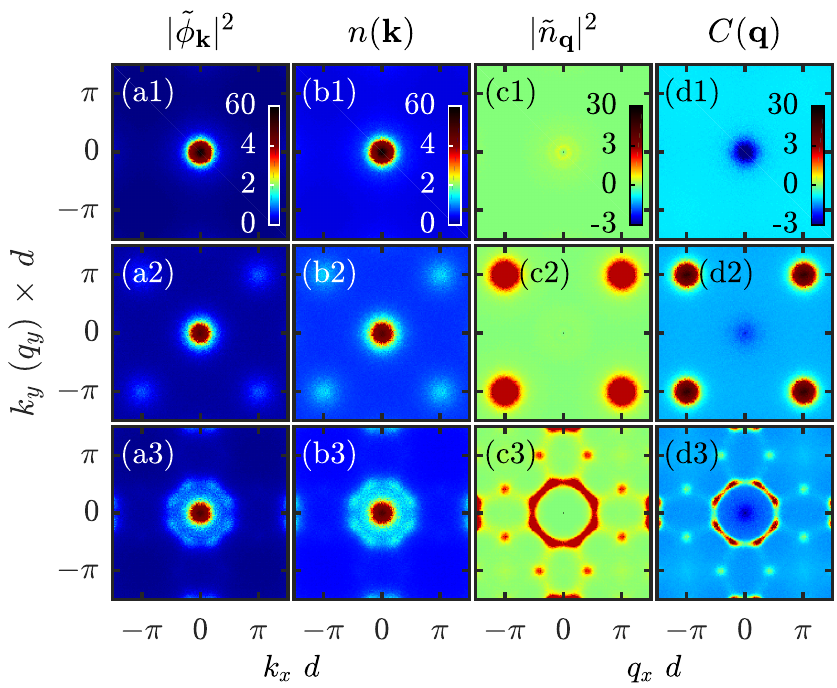} 
	\caption{\textbf{TOF snapshots at time $Ut=28$}. The quench rate is $a_Q/U=0.01$. TOF density $n(\mathbf{k})$ (b1-b3) and covariance $C(\mathbf{q})$ (d1-d3) are compared with the Fourier transform of order parameter $\tilde{\phi}_\mathbf{k}$ (a1-a3) and density $\tilde{n}_\mathbf{q}$ (c1-c3). Box interactions are considered with $r_c/d=\{0$ (non-interacting)$,1,3\}$ and strengths $V_0/U=\{0,0.66,0.36\}$ (top to bottom). }
	\label{fig:Compare_S_TOF}
\end{figure}

When system dynamics is frozen, $\tilde{\phi}_\mathbf{k}$ and $n(\mathbf{k})$ are almost uniform except for a small peak area centred at $\mathbf{k}=0$ (see  Fig.~\ref{fig:Compare_n_C_early}). Such a feature is hardly visible in the distribution of $\tilde{n}_\mathbf{q}$ and $C(\mathbf{q})$. Note that all these distributions are flat in the initial state. At a later time and for weak long-range interactions, $n(k)$ has a higher peak around $k=0$, signifying the appearance of the SF component. Here only the first Brillouin zone is plotted. When the soft-core radius is small, widths of the momentum distribution decreases slowly with increasing interaction strengths (Fig.~\ref{fig:Compare_n_C}a1). Here the roton instability is found  at a large momentum $\sqrt{2}\pi/r_c$ (factor $\sqrt{2}$ is due to the 2D square lattice, see Fig.~\ref{fig:Compare_S_TOF} for details). However, the roton excitation is weak and only perturbs the momentum distribution. When the radius is large, $n(\mathbf{k})$ is affected by rotons (see Fig.~\ref{fig:Compare_n_C}b3). The appearance of the roton minima causes a flat dispersion relation (Fig.~\ref{fig:Compare_n_C}a3). 

To reveal details of the data shown in Fig.~\ref{fig:Compare_n_C}, we calculate $\tilde{\phi}_\mathbf{k}$ and  $\tilde{n}_\mathbf{q}$, i.e.  Fourier transformation of the order parameter and spatial density. Distributions of these two quantities together with $n(\mathbf{k})$ and $C(\mathbf{q})$ are shown in Fig.~\ref{fig:Compare_S_TOF}. In the SF regime, the distributions are more visible in $\tilde{\phi}_\mathbf{k}$ and $n(\mathbf{k})$ where a peak is found around $|\mathbf{k}|=0$ (the first row in Fig.~\ref{fig:Compare_S_TOF}). For stronger interactions, peaks at different momentum are found (the second and third rows in Fig.~\ref{fig:Compare_S_TOF}). These peaks are more profound in the Fourier transformation of the density and covariance. For example, four peaks are found at $|k_x|d=|k_y|d=\pi$ in the second row. This is because these positions are determined by the soft-core radius~\cite{Li2012}, given by $\pi d/r_c$. When $r_c=d$, we thus find the peak positions at $|k_x|d=|k_y|d=\pi$. Increasing the soft-core radius, more and more peaks emerge. An example with $r_c=3d$ is shown  in Fig.~\ref{fig:Compare_S_TOF}a3-d3.  Here the longer-range interactions excite density waves with different characteristic wave-lengths. These extra length scales cause peaks in the momentum distribution.

\section{Quench dynamics with soft-core interaction}
\begin{figure}
	\centering
	\includegraphics[width=\linewidth]{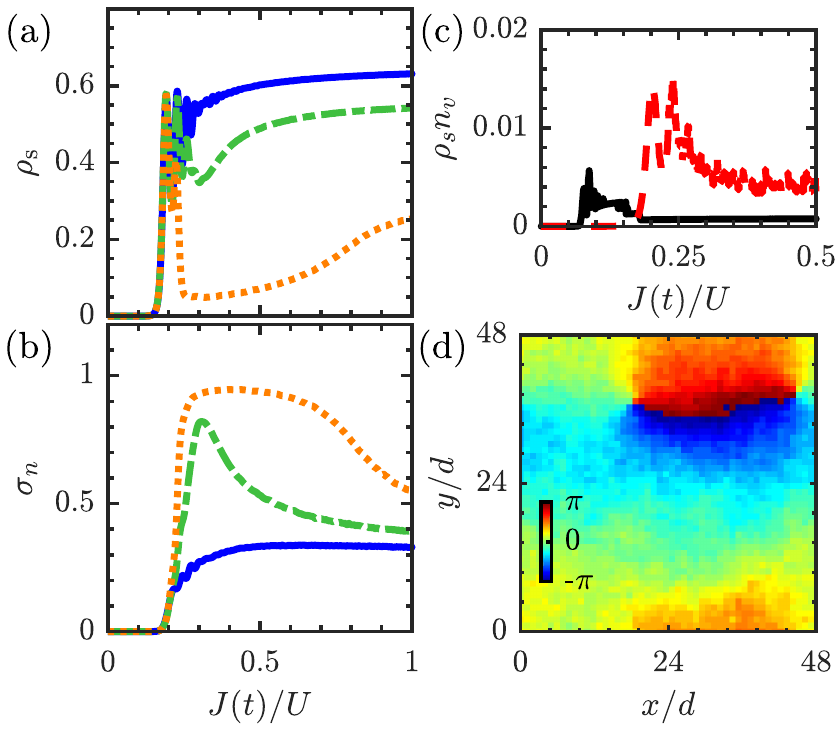} 
	\caption{\textbf{Superfluid fraction (a), density variance (b), and vortex nucleation (c-d) with soft-core interactions.} (a-b), The soft-core radius is $r_c/d=2.5$ and quench rate $a_Q/U=0.01$. The interaction strengths are $V_0/U=0.1$ (blue), $0.5$ (green) and $0.8$ (yellow). (c) Superfluid vortex density under different quench rate $a_Q/U= 10^{-2} {\rm (red)}, 10^{-3} {\rm (black)} $. The interaction strength is $V_0/U=0.5$ and radius $r_c/d=2.5$. (d) The phase of local SF order parameter $\arg(\phi)$ at $J(t_f)=0.5U$ with $a_Q/U=10^{-3}$. } 
	\label{fig:softcore2}
\end{figure}

The more realistic soft-core interaction has a similar shape as the box potential at short distances, while decaying quickly with increasing distances. With the same initial state, the dynamics is qualitatively the same as that of the box interaction. This is demonstrated with an example where $r_c/d = 2.5$. The variables $\rho_s$ and $\sigma_n$ are shown in Fig.~\ref{fig:softcore2}a,b. When the soft-core interaction is weak, the dynamics is again similar to the non-interacting case. But as $V_0$ grows, $\rho_s$ will decrease and $\sigma_n$ increases. The rough border between weak and strong interactions can be estimated by the roton instability, which leads to $J \approx0.29U$. The evolution of the vortex density with varying quench rate is shown in Fig.~\ref{fig:softcore2}c. Compared to Fig.~\ref{fig:vortices}, the notable difference is that decay of $n_v$ becomes relatively faster as $a_Q$ decreases. In the SF region,  the vortex is almost entirely eliminated (Fig.~\ref{fig:softcore2}d). 

\begin{figure}
	\centering
	\includegraphics[width=\linewidth]{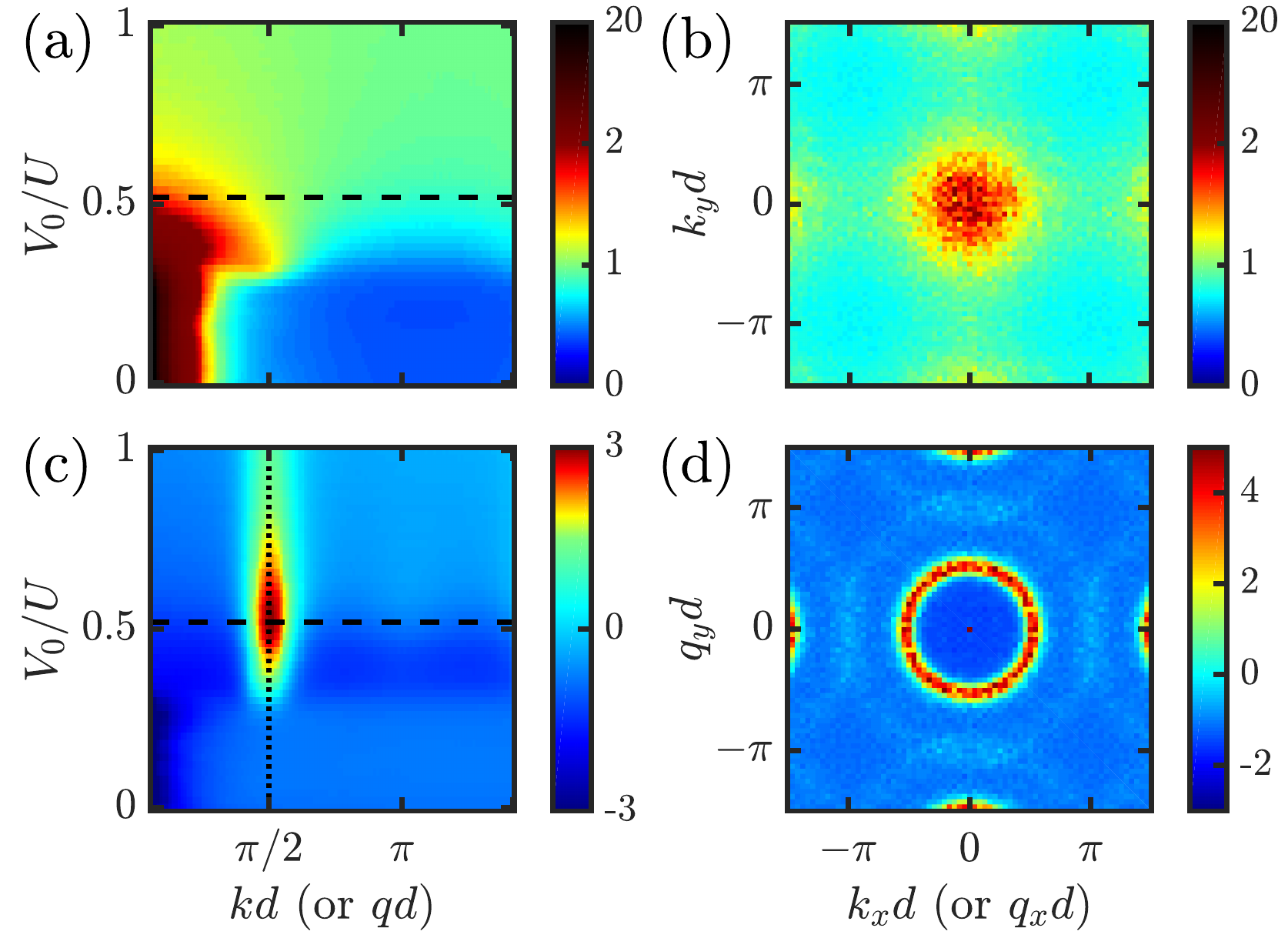} 
	\caption{\textbf{Momentum and covariance for atoms with soft-core interactions.} Radial distribution (a) and two-dimensional distribution (b) of the momentum density $n(\mathbf{k})$. Radial distribution (c) and two-dimensional distribution (d) of the covariance $C(q)$. The soft-core radius $r_c=3d$, quench rate $a_Q/U=0.01$, probe time $Ut=28$, and interaction strength $V_0/U=0.36$. The dotted vertical line marks the momentum corresponding to the roton minimum.  }
	\label{fig:softcore}
\end{figure}

Although dynamics between the box and soft-core interaction is similar, the tail of the soft-core interaction alters the time-of-flight results. As shown in Fig.~\ref{fig:softcore}a,c, the strength that facilitates density structures is similar to Fig.~\ref{fig:Compare_n_C}b2,b3, and the peak positions of $C(\mathbf{q})$ also agree with the roton instability analysis. However, the patterns of covariance are apparently different. For the box interactions, $C(\mathbf{q})$ has a clearly octupole geometry. In the soft-core case, more wave numbers are excited in the dynamics  such that the peak has less contrast at momentum $|\mathbf{q}|=|\mathbf{k}_{\rm rot}|$ (see Fig.~\ref{fig:softcore}d). 

\section{Conclusion}
In this paper, we studied dynamical properties of a two-dimensional eBHM by quenching the atomic hopping. When quenching the hopping from a MI to a SF phase, two stages are found in the dynamical evolution. The dynamics is universal and frozen  initially due to the Kibble-Zurek mechanism. After that, the SF order parameter rises quickly with time (hopping strength). For weaker interactions, we observe universal dynamics even after the onset of the SF order parameter. When the long-range interaction is strong, the universal dynamics disappears due to non-negligible SS components. The system eventually enters the SF regime when the hopping is large enough. We also proposed TOF experiments to measure the quench dynamics and determine effects induced by the long-range interaction. 

In the future, it is worth exploring situations that are relevant to current cold atom experiments. One could investigate roles played by dimensionality (1D, 3D) and lattice structures (triangular, honeycomb, etc) in the dynamics of the Rydberg dressed gases. Another interesting research topic is to explore the quench dynamics induced by time-dependent long-range interactions. This can be realized experimentally by (adiabatically or abruptly) turning on the Rydberg dressing laser~\cite{Zeiher2016}. Starting from a SF state, this permits us to understand, e.g. dynamics of quantum depletion in the presence of long-range interactions. Moreover, it has been shown that ground-state phase transitions in spin models can be probed in the quench dynamics~\cite{Heyl_2013, Paraj_2019}. An interesting question here is whether one could probe the phase transitions (i.e. SS-SF phase transition) in the extended Bose-Hubbard model through monitoring dynamical quantities in a quench process. 

\acknowledgments We thank Huaizhi Wu and Mark Fromhold for many useful comments. This work is supported by the UKIERI-UGC Thematic Partnership No. IND/CONT/G/16-17/73, EPSRC Grants No. EP/M014266/1 and No. EP/R04340X/1, Indo-French Centre for the Promotion of Advanced Research, and National Natural Science Foundation of China under Grants No. 11304386 and No. 11774428. The numerical simulation is carried out at the University of Nottingham High Performance Computing Facility. We also acknowledge the support of the University of Nottingham through a RPA grant and China Scholarship Council. 

\appendix

\section{Time-of-flight interference and noise correlation}

In the time-of-flight experiments, the atomic motion is nearly governed by ballistic expansion. Therefore, the field operator at space-time $\{t;\mathbf{x}\}$ yields $\hat{b}_\mathbf{k} = \tilde{w}(\mathbf{k})\sum_i \hat{b}_i\exp[i\mathbf{k}\cdot\mathbf{r}_i]$ with $\mathbf{k}=m\mathbf{x}/(\hbar t)$. Here $\tilde{w}(\mathbf{k})$ is the Fourier component of Wannier function of the lowest band of the optical lattice. We make an assumption that the optical lattice is deep enough such that $|\tilde{w}(\mathbf{k})|$ is approximately a constant. Under Gutzwiller approximation, the expectation value of density operator $n(\mathbf{k})=\langle\hat{n}_\mathbf{k}\rangle = \langle \hat{b}^\dag_\mathbf{k} \hat{b}_\mathbf{k} \rangle$ reads
\begin{equation} \label{appeq:nk}
n(\mathbf{k}) = | \tilde{\phi}_{\mathbf{k}} |^2 + \frac{1}{M} \sum_i \left(n_i - |\phi_i|^2\right).
\end{equation}
Compared to Eq.~\ref{eq:nk}, the second term in Eq.~\eqref{appeq:nk} is a constant, which preserves the total number of atoms. 

Similarly, the expression for the covariance $C(\mathbf{k},\mathbf{k'}) = \langle \hat{n}_\mathbf{k} \hat{n}_\mathbf{k'} \rangle - \langle \hat{n}_\mathbf{k} \rangle \langle \hat{n}_\mathbf{k'} \rangle$ is found to be
\begin{align} \label{appeq:Ck}
C(\mathbf{k},\mathbf{k}')
	&= \left| F(n-|\phi|^2;\mathbf{q}) \right|^2 + \left|F(\eta^*-\phi^{*2};\mathbf{Q})\right|^2 + \nonumber \\
	&\phantom{{}=} \Delta(\mathbf{q})F(n-|\phi|^2;\mathbf{q}) - 4X(\phi^*;(n-|\phi|^2)\phi) + \nonumber \\
	&\phantom{{}=}  2X(\phi^*;\gamma-\eta\phi^*) + \sqrt{M} \Delta(\mathbf{q})F(\phi^*;\mathbf{k})\tilde{F}(\phi;\mathbf{k}') \nonumber \\
	&\phantom{{}=} + 2 \sqrt{M} \Big[ Y(\phi,\phi,\eta^*-\phi^{*2};\mathbf{k},\mathbf{k}') + \nonumber \\
	&\phantom{{}=} \quad Y(n-|\phi|^2,\phi,\phi^*;\mathbf{q},\mathbf{k}) \Big] + \frac{C_0}{M},
\end{align} 
where $\mathbf{q}=\mathbf{k}'-\mathbf{k}$, $\mathbf{Q}=\mathbf{k}'+\mathbf{k}$, and $\eta_i=\langle \hat{b}_i\hat{b}_i \rangle$, $\gamma_i=\langle \hat{b}^\dag_i\hat{b}_i\hat{b}_i \rangle$, $\lambda_i=\langle \hat{b}^\dag_i\hat{b}^\dag_i\hat{b}_i\hat{b}_i \rangle$. $F$ and $\tilde{F}$ represents for the Fourier and inverse Fourier transformation, and $\Delta(\mathbf{q}) = 1/\sqrt{M}\sum_i \exp[i\mathbf{q}\cdot\mathbf{r}_i]$, $X(f,g) = {\rm Re} [ F(f;\mathbf{k})\tilde{F}(g;\mathbf{k}) + F(f;\mathbf{k}')\tilde{F}(g;\mathbf{k}') ]$, $Y(f,g,h;\mathbf{k}_1,\mathbf{k}_2) = {\rm Re}[F(f;\mathbf{k}_1) F(g;\mathbf{k}_2) \tilde{F}(h;\mathbf{k}_1+\mathbf{k}_2)]$, $C_0 = \sum_i [8n_i|\phi_i|^2 - 2n_i^2 - |\eta_i|^2 + \lambda_i - 6|\phi_i|^4 - 4 {\rm Re} (\gamma^*_i\phi_i - \eta^*\phi_i^2 ) ]$. 

%\bibliographystyle{apsrev4-1}
%\bibliography{references}

\begin{thebibliography}{98}%
	\makeatletter
	\providecommand \@ifxundefined [1]{%
		\@ifx{#1\undefined}
	}%
	\providecommand \@ifnum [1]{%
		\ifnum #1\expandafter \@firstoftwo
		\else \expandafter \@secondoftwo
		\fi
	}%
	\providecommand \@ifx [1]{%
		\ifx #1\expandafter \@firstoftwo
		\else \expandafter \@secondoftwo
		\fi
	}%
	\providecommand \natexlab [1]{#1}%
	\providecommand \enquote  [1]{``#1''}%
	\providecommand \bibnamefont  [1]{#1}%
	\providecommand \bibfnamefont [1]{#1}%
	\providecommand \citenamefont [1]{#1}%
	\providecommand \href@noop [0]{\@secondoftwo}%
	\providecommand \href [0]{\begingroup \@sanitize@url \@href}%
	\providecommand \@href[1]{\@@startlink{#1}\@@href}%
	\providecommand \@@href[1]{\endgroup#1\@@endlink}%
	\providecommand \@sanitize@url [0]{\catcode `\\12\catcode `\$12\catcode
		`\&12\catcode `\#12\catcode `\^12\catcode `\_12\catcode `\%12\relax}%
	\providecommand \@@startlink[1]{}%
	\providecommand \@@endlink[0]{}%
	\providecommand \url  [0]{\begingroup\@sanitize@url \@url }%
	\providecommand \@url [1]{\endgroup\@href {#1}{\urlprefix }}%
	\providecommand \urlprefix  [0]{URL }%
	\providecommand \Eprint [0]{\href }%
	\providecommand \doibase [0]{http://dx.doi.org/}%
	\providecommand \selectlanguage [0]{\@gobble}%
	\providecommand \bibinfo  [0]{\@secondoftwo}%
	\providecommand \bibfield  [0]{\@secondoftwo}%
	\providecommand \translation [1]{[#1]}%
	\providecommand \BibitemOpen [0]{}%
	\providecommand \bibitemStop [0]{}%
	\providecommand \bibitemNoStop [0]{.\EOS\space}%
	\providecommand \EOS [0]{\spacefactor3000\relax}%
	\providecommand \BibitemShut  [1]{\csname bibitem#1\endcsname}%
	\let\auto@bib@innerbib\@empty
	%</preamble>
	\bibitem [{\citenamefont {Jaksch}\ \emph {et~al.}(1998)\citenamefont {Jaksch},
		\citenamefont {Bruder}, \citenamefont {Cirac}, \citenamefont {Gardiner},\
		and\ \citenamefont {Zoller}}]{Jaksch1998}%
	\BibitemOpen
	\bibfield  {author} {\bibinfo {author} {\bibfnamefont {D.}~\bibnamefont
			{Jaksch}}, \bibinfo {author} {\bibfnamefont {C.}~\bibnamefont {Bruder}},
		\bibinfo {author} {\bibfnamefont {J.~I.}\ \bibnamefont {Cirac}}, \bibinfo
		{author} {\bibfnamefont {C.~W.}\ \bibnamefont {Gardiner}}, \ and\ \bibinfo
		{author} {\bibfnamefont {P.}~\bibnamefont {Zoller}},\ }\href {\doibase
		10.1103/PhysRevLett.81.3108} {\bibfield  {journal} {\bibinfo  {journal}
			{Phys. Rev. Lett.}\ }\textbf {\bibinfo {volume} {81}},\ \bibinfo {pages}
		{3108} (\bibinfo {year} {1998})}\BibitemShut {NoStop}%
	\bibitem [{\citenamefont {Stenger}\ \emph {et~al.}(1999)\citenamefont
		{Stenger}, \citenamefont {Inouye}, \citenamefont {Andrews}, \citenamefont
		{Miesner}, \citenamefont {Stamper-Kurn},\ and\ \citenamefont
		{Ketterle}}]{Stenger1999}%
	\BibitemOpen
	\bibfield  {author} {\bibinfo {author} {\bibfnamefont {J.}~\bibnamefont
			{Stenger}}, \bibinfo {author} {\bibfnamefont {S.}~\bibnamefont {Inouye}},
		\bibinfo {author} {\bibfnamefont {M.~R.}\ \bibnamefont {Andrews}}, \bibinfo
		{author} {\bibfnamefont {H.~J.}\ \bibnamefont {Miesner}}, \bibinfo {author}
		{\bibfnamefont {D.~M.}\ \bibnamefont {Stamper-Kurn}}, \ and\ \bibinfo
		{author} {\bibfnamefont {W.}~\bibnamefont {Ketterle}},\ }\href {\doibase
		10.1103/PhysRevLett.82.2422} {\bibfield  {journal} {\bibinfo  {journal}
			{Phys. Rev. Lett.}\ }\textbf {\bibinfo {volume} {82}},\ \bibinfo {pages}
		{2422} (\bibinfo {year} {1999})}\BibitemShut {NoStop}%
	\bibitem [{\citenamefont {Theis}\ \emph {et~al.}(2004)\citenamefont {Theis},
		\citenamefont {Thalhammer}, \citenamefont {Winkler}, \citenamefont {Hellwig},
		\citenamefont {Ruff}, \citenamefont {Grimm},\ and\ \citenamefont
		{Denschlag}}]{Theis2004}%
	\BibitemOpen
	\bibfield  {author} {\bibinfo {author} {\bibfnamefont {M.}~\bibnamefont
			{Theis}}, \bibinfo {author} {\bibfnamefont {G.}~\bibnamefont {Thalhammer}},
		\bibinfo {author} {\bibfnamefont {K.}~\bibnamefont {Winkler}}, \bibinfo
		{author} {\bibfnamefont {M.}~\bibnamefont {Hellwig}}, \bibinfo {author}
		{\bibfnamefont {G.}~\bibnamefont {Ruff}}, \bibinfo {author} {\bibfnamefont
			{R.}~\bibnamefont {Grimm}}, \ and\ \bibinfo {author} {\bibfnamefont {J.~H.}\
			\bibnamefont {Denschlag}},\ }\href {\doibase 10.1103/PhysRevLett.93.123001}
	{\bibfield  {journal} {\bibinfo  {journal} {Phys. Rev. Lett.}\ }\textbf
		{\bibinfo {volume} {93}},\ \bibinfo {pages} {123001} (\bibinfo {year}
		{2004})}\BibitemShut {NoStop}%
	\bibitem [{\citenamefont {Chin}\ \emph {et~al.}(2010)\citenamefont {Chin},
		\citenamefont {Grimm}, \citenamefont {Julienne},\ and\ \citenamefont
		{Tiesinga}}]{Chin2010}%
	\BibitemOpen
	\bibfield  {author} {\bibinfo {author} {\bibfnamefont {C.}~\bibnamefont
			{Chin}}, \bibinfo {author} {\bibfnamefont {R.}~\bibnamefont {Grimm}},
		\bibinfo {author} {\bibfnamefont {P.}~\bibnamefont {Julienne}}, \ and\
		\bibinfo {author} {\bibfnamefont {E.}~\bibnamefont {Tiesinga}},\ }\href
	{\doibase 10.1103/RevModPhys.82.1225} {\bibfield  {journal} {\bibinfo
			{journal} {Rev. Mod. Phys.}\ }\textbf {\bibinfo {volume} {82}},\ \bibinfo
		{pages} {1225} (\bibinfo {year} {2010})}\BibitemShut {NoStop}%
	\bibitem [{\citenamefont {Schachenmayer}\ \emph {et~al.}(2010)\citenamefont
		{Schachenmayer}, \citenamefont {Lesanovsky}, \citenamefont {Micheli},\ and\
		\citenamefont {Daley}}]{Schachenmayer2010}%
	\BibitemOpen
	\bibfield  {author} {\bibinfo {author} {\bibfnamefont {J.}~\bibnamefont
			{Schachenmayer}}, \bibinfo {author} {\bibfnamefont {I.}~\bibnamefont
			{Lesanovsky}}, \bibinfo {author} {\bibfnamefont {A.}~\bibnamefont {Micheli}},
		\ and\ \bibinfo {author} {\bibfnamefont {A.~J.}\ \bibnamefont {Daley}},\
	}\href {\doibase 10.1088/1367-2630/12/10/103044} {\bibfield  {journal}
		{\bibinfo  {journal} {New J. Phys.}\ }\textbf {\bibinfo {volume} {12}},\
		\bibinfo {pages} {103044} (\bibinfo {year} {2010})}\BibitemShut {NoStop}%
	\bibitem [{\citenamefont {Johnson}\ and\ \citenamefont
		{Rolston}(2010)}]{Johnson2010}%
	\BibitemOpen
	\bibfield  {author} {\bibinfo {author} {\bibfnamefont {J.~E.}\ \bibnamefont
			{Johnson}}\ and\ \bibinfo {author} {\bibfnamefont {S.~L.}\ \bibnamefont
			{Rolston}},\ }\href {\doibase 10.1103/PhysRevA.82.033412} {\bibfield
		{journal} {\bibinfo  {journal} {Phys. Rev. A}\ }\textbf {\bibinfo {volume}
			{82}},\ \bibinfo {pages} {033412} (\bibinfo {year} {2010})}\BibitemShut
	{NoStop}%
	\bibitem [{\citenamefont {Anderson}\ \emph {et~al.}(2011)\citenamefont
		{Anderson}, \citenamefont {Younge},\ and\ \citenamefont
		{Raithel}}]{Anderson2011}%
	\BibitemOpen
	\bibfield  {author} {\bibinfo {author} {\bibfnamefont {S.~E.}\ \bibnamefont
			{Anderson}}, \bibinfo {author} {\bibfnamefont {K.~C.}\ \bibnamefont
			{Younge}}, \ and\ \bibinfo {author} {\bibfnamefont {G.}~\bibnamefont
			{Raithel}},\ }\href {\doibase 10.1103/PhysRevLett.107.263001} {\bibfield
		{journal} {\bibinfo  {journal} {Phys. Rev. Lett.}\ }\textbf {\bibinfo
			{volume} {107}},\ \bibinfo {pages} {263001} (\bibinfo {year}
		{2011})}\BibitemShut {NoStop}%
	\bibitem [{\citenamefont {Viteau}\ \emph {et~al.}(2011)\citenamefont {Viteau},
		\citenamefont {Bason}, \citenamefont {Radogostowicz}, \citenamefont
		{Malossi}, \citenamefont {Ciampini}, \citenamefont {Morsch},\ and\
		\citenamefont {Arimondo}}]{Viteau2011}%
	\BibitemOpen
	\bibfield  {author} {\bibinfo {author} {\bibfnamefont {M.}~\bibnamefont
			{Viteau}}, \bibinfo {author} {\bibfnamefont {M.~G.}\ \bibnamefont {Bason}},
		\bibinfo {author} {\bibfnamefont {J.}~\bibnamefont {Radogostowicz}}, \bibinfo
		{author} {\bibfnamefont {N.}~\bibnamefont {Malossi}}, \bibinfo {author}
		{\bibfnamefont {D.}~\bibnamefont {Ciampini}}, \bibinfo {author}
		{\bibfnamefont {O.}~\bibnamefont {Morsch}}, \ and\ \bibinfo {author}
		{\bibfnamefont {E.}~\bibnamefont {Arimondo}},\ }\href {\doibase
		10.1103/PhysRevLett.107.060402} {\bibfield  {journal} {\bibinfo  {journal}
			{Phys. Rev. Lett.}\ }\textbf {\bibinfo {volume} {107}},\ \bibinfo {pages}
		{060402} (\bibinfo {year} {2011})}\BibitemShut {NoStop}%
	\bibitem [{\citenamefont {Macr{\`{i}}}\ and\ \citenamefont
		{Pohl}(2014)}]{Macri2014}%
	\BibitemOpen
	\bibfield  {author} {\bibinfo {author} {\bibfnamefont {T.}~\bibnamefont
			{Macr{\`{i}}}}\ and\ \bibinfo {author} {\bibfnamefont {T.}~\bibnamefont
			{Pohl}},\ }\href {\doibase 10.1103/PhysRevA.89.011402} {\bibfield  {journal}
		{\bibinfo  {journal} {Phys. Rev. A}\ }\textbf {\bibinfo {volume} {89}},\
		\bibinfo {pages} {011402(R)} (\bibinfo {year} {2014})}\BibitemShut {NoStop}%
	\bibitem [{\citenamefont {Gross}\ and\ \citenamefont
		{Bloch}(2017)}]{Gross2017}%
	\BibitemOpen
	\bibfield  {author} {\bibinfo {author} {\bibfnamefont {C.}~\bibnamefont
			{Gross}}\ and\ \bibinfo {author} {\bibfnamefont {I.}~\bibnamefont {Bloch}},\
	}\href {\doibase 10.1126/science.aal3837} {\bibfield  {journal} {\bibinfo
			{journal} {Science}\ }\textbf {\bibinfo {volume} {357}},\ \bibinfo {pages}
		{995} (\bibinfo {year} {2017})}\BibitemShut {NoStop}%
	\bibitem [{\citenamefont {Fisher}\ \emph {et~al.}(1989)\citenamefont {Fisher},
		\citenamefont {Weichman}, \citenamefont {Grinstein},\ and\ \citenamefont
		{Fisher}}]{Fisher1989}%
	\BibitemOpen
	\bibfield  {author} {\bibinfo {author} {\bibfnamefont {M.~P.~A.}\ \bibnamefont
			{Fisher}}, \bibinfo {author} {\bibfnamefont {P.~B.}\ \bibnamefont
			{Weichman}}, \bibinfo {author} {\bibfnamefont {G.}~\bibnamefont {Grinstein}},
		\ and\ \bibinfo {author} {\bibfnamefont {D.~S.}\ \bibnamefont {Fisher}},\
	}\href {\doibase 10.1103/PhysRevB.40.546} {\bibfield  {journal} {\bibinfo
			{journal} {Phys. Rev. B}\ }\textbf {\bibinfo {volume} {40}},\ \bibinfo
		{pages} {546} (\bibinfo {year} {1989})}\BibitemShut {NoStop}%
	\bibitem [{\citenamefont {Greiner}\ \emph
		{et~al.}(2002{\natexlab{a}})\citenamefont {Greiner}, \citenamefont {Mandel},
		\citenamefont {Esslinger}, \citenamefont {H{\"{a}}nsch},\ and\ \citenamefont
		{Bloch}}]{Greiner2002a}%
	\BibitemOpen
	\bibfield  {author} {\bibinfo {author} {\bibfnamefont {M.}~\bibnamefont
			{Greiner}}, \bibinfo {author} {\bibfnamefont {O.}~\bibnamefont {Mandel}},
		\bibinfo {author} {\bibfnamefont {T.}~\bibnamefont {Esslinger}}, \bibinfo
		{author} {\bibfnamefont {T.~W.}\ \bibnamefont {H{\"{a}}nsch}}, \ and\
		\bibinfo {author} {\bibfnamefont {I.}~\bibnamefont {Bloch}},\ }\href
	{\doibase 10.1038/415039a} {\bibfield  {journal} {\bibinfo  {journal}
			{Nature (London)}\ }\textbf {\bibinfo {volume} {415}},\ \bibinfo {pages} {39}
		(\bibinfo {year} {2002}{\natexlab{a}})}\BibitemShut {NoStop}%
	\bibitem [{\citenamefont {Jaksch}\ and\ \citenamefont
		{Zoller}(2005)}]{Jaksch2005}%
	\BibitemOpen
	\bibfield  {author} {\bibinfo {author} {\bibfnamefont {D.}~\bibnamefont
			{Jaksch}}\ and\ \bibinfo {author} {\bibfnamefont {P.}~\bibnamefont
			{Zoller}},\ }\href {\doibase 10.1016/j.aop.2004.09.010} {\bibfield  {journal}
		{\bibinfo  {journal} {Ann. Phys. (N. Y.)}\ }\textbf {\bibinfo {volume}
			{315}},\ \bibinfo {pages} {52} (\bibinfo {year} {2005})}\BibitemShut
	{NoStop}%
	\bibitem [{\citenamefont {Lewenstein}\ \emph {et~al.}(2007)\citenamefont
		{Lewenstein}, \citenamefont {Sanpera}, \citenamefont {Ahufinger},
		\citenamefont {Damski}, \citenamefont {Sen},\ and\ \citenamefont
		{Sen}}]{Lewenstein2007}%
	\BibitemOpen
	\bibfield  {author} {\bibinfo {author} {\bibfnamefont {M.}~\bibnamefont
			{Lewenstein}}, \bibinfo {author} {\bibfnamefont {A.}~\bibnamefont {Sanpera}},
		\bibinfo {author} {\bibfnamefont {V.}~\bibnamefont {Ahufinger}}, \bibinfo
		{author} {\bibfnamefont {B.}~\bibnamefont {Damski}}, \bibinfo {author}
		{\bibfnamefont {A.}~\bibnamefont {Sen}}, \ and\ \bibinfo {author}
		{\bibfnamefont {U.}~\bibnamefont {Sen}},\ }\href {\doibase
		10.1080/00018730701223200} {\bibfield  {journal} {\bibinfo  {journal} {Adv.
				Phys.}\ }\textbf {\bibinfo {volume} {56}},\ \bibinfo {pages} {243} (\bibinfo
		{year} {2007})}\BibitemShut {NoStop}%
	\bibitem [{\citenamefont {Bloch}\ \emph {et~al.}(2008)\citenamefont {Bloch},
		\citenamefont {Dalibard},\ and\ \citenamefont {Zwerger}}]{Bloch2008}%
	\BibitemOpen
	\bibfield  {author} {\bibinfo {author} {\bibfnamefont {I.}~\bibnamefont
			{Bloch}}, \bibinfo {author} {\bibfnamefont {J.}~\bibnamefont {Dalibard}}, \
		and\ \bibinfo {author} {\bibfnamefont {W.}~\bibnamefont {Zwerger}},\ }\href
	{\doibase 10.1103/RevModPhys.80.885} {\bibfield  {journal} {\bibinfo
			{journal} {Rev. Mod. Phys.}\ }\textbf {\bibinfo {volume} {80}},\ \bibinfo
		{pages} {885} (\bibinfo {year} {2008})}\BibitemShut {NoStop}%
	\bibitem [{\citenamefont {van Oosten}\ \emph {et~al.}(2001)\citenamefont {van
			Oosten}, \citenamefont {van~der Straten},\ and\ \citenamefont
		{Stoof}}]{VanOosten2001}%
	\BibitemOpen
	\bibfield  {author} {\bibinfo {author} {\bibfnamefont {D.}~\bibnamefont {van
				Oosten}}, \bibinfo {author} {\bibfnamefont {P.}~\bibnamefont {van~der
				Straten}}, \ and\ \bibinfo {author} {\bibfnamefont {H.~T.~C.}\ \bibnamefont
			{Stoof}},\ }\href {\doibase 10.1103/PhysRevA.63.053601} {\bibfield  {journal}
		{\bibinfo  {journal} {Phys. Rev. A}\ }\textbf {\bibinfo {volume} {63}},\
		\bibinfo {pages} {053601} (\bibinfo {year} {2001})}\BibitemShut {NoStop}%
	\bibitem [{\citenamefont {Zwerger}(2003)}]{Zwerger2003}%
	\BibitemOpen
	\bibfield  {author} {\bibinfo {author} {\bibfnamefont {W.}~\bibnamefont
			{Zwerger}},\ }\href {\doibase 10.1088/1464-4266/5/2/352} {\bibfield
		{journal} {\bibinfo  {journal} {J. Opt. B}\ }\textbf {\bibinfo {volume}
			{5}},\ \bibinfo {pages} {S9} (\bibinfo {year} {2003})}\BibitemShut {NoStop}%
	\bibitem [{\citenamefont {Wall}\ \emph {et~al.}(2016)\citenamefont {Wall},
		\citenamefont {Koller}, \citenamefont {Li}, \citenamefont {Zhang},
		\citenamefont {Cooper}, \citenamefont {Ye},\ and\ \citenamefont
		{Rey}}]{Wall2016}%
	\BibitemOpen
	\bibfield  {author} {\bibinfo {author} {\bibfnamefont {M.~L.}\ \bibnamefont
			{Wall}}, \bibinfo {author} {\bibfnamefont {A.~P.}\ \bibnamefont {Koller}},
		\bibinfo {author} {\bibfnamefont {S.}~\bibnamefont {Li}}, \bibinfo {author}
		{\bibfnamefont {X.}~\bibnamefont {Zhang}}, \bibinfo {author} {\bibfnamefont
			{N.~R.}\ \bibnamefont {Cooper}}, \bibinfo {author} {\bibfnamefont
			{J.}~\bibnamefont {Ye}}, \ and\ \bibinfo {author} {\bibfnamefont {A.~M.}\
			\bibnamefont {Rey}},\ }\href {\doibase 10.1103/PhysRevLett.116.035301}
	{\bibfield  {journal} {\bibinfo  {journal} {Phys. Rev. Lett.}\ }\textbf
		{\bibinfo {volume} {116}},\ \bibinfo {pages} {035301} (\bibinfo {year}
		{2016})}\BibitemShut {NoStop}%
	\bibitem [{\citenamefont {Kolkowitz}\ \emph {et~al.}(2017)\citenamefont
		{Kolkowitz}, \citenamefont {Bromley}, \citenamefont {Bothwell}, \citenamefont
		{Wall}, \citenamefont {Marti}, \citenamefont {Koller}, \citenamefont {Zhang},
		\citenamefont {Rey},\ and\ \citenamefont {Ye}}]{Bromley2016}%
	\BibitemOpen
	\bibfield  {author} {\bibinfo {author} {\bibfnamefont {S.}~\bibnamefont
			{Kolkowitz}}, \bibinfo {author} {\bibfnamefont {S.~L.}\ \bibnamefont
			{Bromley}}, \bibinfo {author} {\bibfnamefont {T.}~\bibnamefont {Bothwell}},
		\bibinfo {author} {\bibfnamefont {M.~L.}\ \bibnamefont {Wall}}, \bibinfo
		{author} {\bibfnamefont {G.~E.}\ \bibnamefont {Marti}}, \bibinfo {author}
		{\bibfnamefont {A.~P.}\ \bibnamefont {Koller}}, \bibinfo {author}
		{\bibfnamefont {X.}~\bibnamefont {Zhang}}, \bibinfo {author} {\bibfnamefont
			{A.~M.}\ \bibnamefont {Rey}}, \ and\ \bibinfo {author} {\bibfnamefont
			{J.}~\bibnamefont {Ye}},\ }\href {\doibase 10.1038/nature20811} {\bibfield
		{journal} {\bibinfo  {journal} {Nature}\ }\textbf {\bibinfo {volume} {542}},\
		\bibinfo {pages} {66} (\bibinfo {year} {2017})}\BibitemShut {NoStop}%
	\bibitem [{\citenamefont {Zhang}\ and\ \citenamefont {Liu}(2018)}]{Zhang2018}%
	\BibitemOpen
	\bibfield  {author} {\bibinfo {author} {\bibfnamefont {L.}~\bibnamefont
			{Zhang}}\ and\ \bibinfo {author} {\bibfnamefont {X.-J.}\ \bibnamefont
			{Liu}},\ }in\ \href {\doibase 10.1142/9789813272538_0001} {\emph {\bibinfo
			{booktitle} {Synthetic Spin-Orbit Coupling Cold Atoms}}}\ (\bibinfo
	{publisher} {WORLD SCIENTIFIC},\ \bibinfo {year} {2018})\ pp.\ \bibinfo
	{pages} {1--87}\BibitemShut {NoStop}%
	\bibitem [{\citenamefont {Sengupta}\ \emph {et~al.}(2005)\citenamefont
		{Sengupta}, \citenamefont {Pryadko}, \citenamefont {Alet}, \citenamefont
		{Troyer},\ and\ \citenamefont {Schmid}}]{Sengupta2005}%
	\BibitemOpen
	\bibfield  {author} {\bibinfo {author} {\bibfnamefont {P.}~\bibnamefont
			{Sengupta}}, \bibinfo {author} {\bibfnamefont {L.~P.}\ \bibnamefont
			{Pryadko}}, \bibinfo {author} {\bibfnamefont {F.}~\bibnamefont {Alet}},
		\bibinfo {author} {\bibfnamefont {M.}~\bibnamefont {Troyer}}, \ and\ \bibinfo
		{author} {\bibfnamefont {G.}~\bibnamefont {Schmid}},\ }\href {\doibase
		10.1103/PhysRevLett.94.207202} {\bibfield  {journal} {\bibinfo  {journal}
			{Phys. Rev. Lett.}\ }\textbf {\bibinfo {volume} {94}},\ \bibinfo {pages}
		{207202} (\bibinfo {year} {2005})}\BibitemShut {NoStop}%
	\bibitem [{\citenamefont {Scarola}\ \emph {et~al.}(2006)\citenamefont
		{Scarola}, \citenamefont {Demler},\ and\ \citenamefont {{Das
				Sarma}}}]{Scarola2006}%
	\BibitemOpen
	\bibfield  {author} {\bibinfo {author} {\bibfnamefont {V.~W.}\ \bibnamefont
			{Scarola}}, \bibinfo {author} {\bibfnamefont {E.}~\bibnamefont {Demler}}, \
		and\ \bibinfo {author} {\bibfnamefont {S.}~\bibnamefont {{Das Sarma}}},\
	}\href {\doibase 10.1103/PhysRevA.73.051601} {\bibfield  {journal} {\bibinfo
			{journal} {Phys. Rev. A}\ }\textbf {\bibinfo {volume} {73}},\ \bibinfo
		{pages} {051601(R)} (\bibinfo {year} {2006})}\BibitemShut {NoStop}%
	\bibitem [{\citenamefont {Batrouni}\ \emph {et~al.}(2006)\citenamefont
		{Batrouni}, \citenamefont {H{\'{e}}bert},\ and\ \citenamefont
		{Scalettar}}]{Batrouni2006}%
	\BibitemOpen
	\bibfield  {author} {\bibinfo {author} {\bibfnamefont {G.~G.}\ \bibnamefont
			{Batrouni}}, \bibinfo {author} {\bibfnamefont {F.}~\bibnamefont
			{H{\'{e}}bert}}, \ and\ \bibinfo {author} {\bibfnamefont {R.~T.}\
			\bibnamefont {Scalettar}},\ }\href {\doibase 10.1103/PhysRevLett.97.087209}
	{\bibfield  {journal} {\bibinfo  {journal} {Phys. Rev. Lett.}\ }\textbf
		{\bibinfo {volume} {97}},\ \bibinfo {pages} {087209} (\bibinfo {year}
		{2006})}\BibitemShut {NoStop}%
	\bibitem [{\citenamefont {Menotti}\ \emph {et~al.}(2007)\citenamefont
		{Menotti}, \citenamefont {Trefzger},\ and\ \citenamefont
		{Lewenstein}}]{Menotti2007}%
	\BibitemOpen
	\bibfield  {author} {\bibinfo {author} {\bibfnamefont {C.}~\bibnamefont
			{Menotti}}, \bibinfo {author} {\bibfnamefont {C.}~\bibnamefont {Trefzger}}, \
		and\ \bibinfo {author} {\bibfnamefont {M.}~\bibnamefont {Lewenstein}},\
	}\href {\doibase 10.1103/PhysRevLett.98.235301} {\bibfield  {journal}
		{\bibinfo  {journal} {Phys. Rev. Lett.}\ }\textbf {\bibinfo {volume} {98}},\
		\bibinfo {pages} {235301} (\bibinfo {year} {2007})}\BibitemShut {NoStop}%
	\bibitem [{\citenamefont {Yi}\ \emph {et~al.}(2007)\citenamefont {Yi},
		\citenamefont {Li},\ and\ \citenamefont {Sun}}]{Yi2007}%
	\BibitemOpen
	\bibfield  {author} {\bibinfo {author} {\bibfnamefont {S.}~\bibnamefont
			{Yi}}, \bibinfo {author} {\bibfnamefont {T.}~\bibnamefont {Li}}, \ and\
		\bibinfo {author} {\bibfnamefont {C.~P.}\ \bibnamefont {Sun}},\ }\href
	{\doibase 10.1103/PhysRevLett.98.260405} {\bibfield  {journal} {\bibinfo
			{journal} {Phys. Rev. Lett.}\ }\textbf {\bibinfo {volume} {98}},\ \bibinfo
		{pages} {260405} (\bibinfo {year} {2007})}\BibitemShut {NoStop}%
	\bibitem [{\citenamefont {Iskin}\ and\ \citenamefont
		{Freericks}(2009)}]{Iskin2009}%
	\BibitemOpen
	\bibfield  {author} {\bibinfo {author} {\bibfnamefont {M.}~\bibnamefont
			{Iskin}}\ and\ \bibinfo {author} {\bibfnamefont {J.~K.}\ \bibnamefont
			{Freericks}},\ }\href {\doibase 10.1103/PhysRevA.79.053634} {\bibfield
		{journal} {\bibinfo  {journal} {Phys. Rev. A}\ }\textbf {\bibinfo {volume}
			{79}},\ \bibinfo {pages} {053634} (\bibinfo {year} {2009})}\BibitemShut
	{NoStop}%
	\bibitem [{\citenamefont {Iskin}(2011)}]{Iskin2011}%
	\BibitemOpen
	\bibfield  {author} {\bibinfo {author} {\bibfnamefont {M.}~\bibnamefont
			{Iskin}},\ }\href {\doibase 10.1103/PhysRevA.83.051606} {\bibfield  {journal}
		{\bibinfo  {journal} {Phys. Rev. A}\ }\textbf {\bibinfo {volume} {83}},\
		\bibinfo {pages} {051606(R)} (\bibinfo {year} {2011})}\BibitemShut {NoStop}%
	\bibitem [{\citenamefont {Li}\ \emph {et~al.}(2012)\citenamefont {Li},
		\citenamefont {Hamadeh},\ and\ \citenamefont {Lesanovsky}}]{Li2012}%
	\BibitemOpen
	\bibfield  {author} {\bibinfo {author} {\bibfnamefont {W.}~\bibnamefont
			{Li}}, \bibinfo {author} {\bibfnamefont {L.}~\bibnamefont {Hamadeh}}, \ and\
		\bibinfo {author} {\bibfnamefont {I.}~\bibnamefont {Lesanovsky}},\ }\href
	{\doibase 10.1103/PhysRevA.85.053615} {\bibfield  {journal} {\bibinfo
			{journal} {Phys. Rev. A}\ }\textbf {\bibinfo {volume} {85}},\ \bibinfo
		{pages} {053615} (\bibinfo {year} {2012})}\BibitemShut {NoStop}%
	\bibitem [{\citenamefont {Landig}\ \emph {et~al.}(2016)\citenamefont {Landig},
		\citenamefont {Hruby}, \citenamefont {Dogra}, \citenamefont {Landini},
		\citenamefont {Mottl}, \citenamefont {Donner},\ and\ \citenamefont
		{Esslinger}}]{Landig2016}%
	\BibitemOpen
	\bibfield  {author} {\bibinfo {author} {\bibfnamefont {R.}~\bibnamefont
			{Landig}}, \bibinfo {author} {\bibfnamefont {L.}~\bibnamefont {Hruby}},
		\bibinfo {author} {\bibfnamefont {N.}~\bibnamefont {Dogra}}, \bibinfo
		{author} {\bibfnamefont {M.}~\bibnamefont {Landini}}, \bibinfo {author}
		{\bibfnamefont {R.}~\bibnamefont {Mottl}}, \bibinfo {author} {\bibfnamefont
			{T.}~\bibnamefont {Donner}}, \ and\ \bibinfo {author} {\bibfnamefont
			{T.}~\bibnamefont {Esslinger}},\ }\href {\doibase 10.1038/nature17409}
	{\bibfield  {journal} {\bibinfo  {journal} {Nature}\ }\textbf {\bibinfo
			{volume} {532}},\ \bibinfo {pages} {476} (\bibinfo {year}
		{2016})}\BibitemShut {NoStop}%
	\bibitem [{\citenamefont {Li}\ \emph {et~al.}(2018)\citenamefont {Li},
		\citenamefont {Gei{\ss}ler}, \citenamefont {Hofstetter},\ and\ \citenamefont
		{Li}}]{Li2018}%
	\BibitemOpen
	\bibfield  {author} {\bibinfo {author} {\bibfnamefont {Y.}~\bibnamefont
			{Li}}, \bibinfo {author} {\bibfnamefont {A.}~\bibnamefont {Gei{\ss}ler}},
		\bibinfo {author} {\bibfnamefont {W.}~\bibnamefont {Hofstetter}}, \ and\
		\bibinfo {author} {\bibfnamefont {W.}~\bibnamefont {Li}},\ }\href {\doibase
		10.1103/PhysRevA.97.023619} {\bibfield  {journal} {\bibinfo  {journal} {Phys.
				Rev. A}\ }\textbf {\bibinfo {volume} {97}},\ \bibinfo {pages} {023619}
		(\bibinfo {year} {2018})}\BibitemShut {NoStop}%
	\bibitem [{\citenamefont {Alba}\ \emph {et~al.}(2013)\citenamefont {Alba},
		\citenamefont {Haque},\ and\ \citenamefont {L{\"{a}}uchli}}]{Alba2013}%
	\BibitemOpen
	\bibfield  {author} {\bibinfo {author} {\bibfnamefont {V.}~\bibnamefont
			{Alba}}, \bibinfo {author} {\bibfnamefont {M.}~\bibnamefont {Haque}}, \ and\
		\bibinfo {author} {\bibfnamefont {A.~M.}\ \bibnamefont {L{\"{a}}uchli}},\
	}\href {\doibase 10.1103/PhysRevLett.110.260403} {\bibfield  {journal}
		{\bibinfo  {journal} {Phys. Rev. Lett.}\ }\textbf {\bibinfo {volume} {110}},\
		\bibinfo {pages} {260403} (\bibinfo {year} {2013})}\BibitemShut {NoStop}%
	\bibitem [{\citenamefont {Islam}\ \emph {et~al.}(2015)\citenamefont {Islam},
		\citenamefont {Ma}, \citenamefont {Preiss}, \citenamefont {{Eric Tai}},
		\citenamefont {Lukin}, \citenamefont {Rispoli},\ and\ \citenamefont
		{Greiner}}]{Lukin2015}%
	\BibitemOpen
	\bibfield  {author} {\bibinfo {author} {\bibfnamefont {R.}~\bibnamefont
			{Islam}}, \bibinfo {author} {\bibfnamefont {R.}~\bibnamefont {Ma}}, \bibinfo
		{author} {\bibfnamefont {P.~M.}\ \bibnamefont {Preiss}}, \bibinfo {author}
		{\bibfnamefont {M.}~\bibnamefont {{Eric Tai}}}, \bibinfo {author}
		{\bibfnamefont {A.}~\bibnamefont {Lukin}}, \bibinfo {author} {\bibfnamefont
			{M.}~\bibnamefont {Rispoli}}, \ and\ \bibinfo {author} {\bibfnamefont
			{M.}~\bibnamefont {Greiner}},\ }\href {\doibase 10.1038/nature15750}
	{\bibfield  {journal} {\bibinfo  {journal} {Nature}\ }\textbf {\bibinfo
			{volume} {528}},\ \bibinfo {pages} {77} (\bibinfo {year} {2015})}\BibitemShut
	{NoStop}%
	\bibitem [{\citenamefont {Goldman}\ \emph {et~al.}(2016)\citenamefont
		{Goldman}, \citenamefont {Budich},\ and\ \citenamefont
		{Zoller}}]{Goldman2016}%
	\BibitemOpen
	\bibfield  {author} {\bibinfo {author} {\bibfnamefont {N.}~\bibnamefont
			{Goldman}}, \bibinfo {author} {\bibfnamefont {J.~C.}\ \bibnamefont {Budich}},
		\ and\ \bibinfo {author} {\bibfnamefont {P.}~\bibnamefont {Zoller}},\ }\href
	{\doibase 10.1038/nphys3803} {\bibfield  {journal} {\bibinfo  {journal} {Nat.
				Phys.}\ }\textbf {\bibinfo {volume} {12}},\ \bibinfo {pages} {639} (\bibinfo
		{year} {2016})}\BibitemShut {NoStop}%
	\bibitem [{\citenamefont {Lohse}\ \emph {et~al.}(2016)\citenamefont {Lohse},
		\citenamefont {Schweizer}, \citenamefont {Zilberberg}, \citenamefont
		{Aidelsburger},\ and\ \citenamefont {Bloch}}]{Lohse2016}%
	\BibitemOpen
	\bibfield  {author} {\bibinfo {author} {\bibfnamefont {M.}~\bibnamefont
			{Lohse}}, \bibinfo {author} {\bibfnamefont {C.}~\bibnamefont {Schweizer}},
		\bibinfo {author} {\bibfnamefont {O.}~\bibnamefont {Zilberberg}}, \bibinfo
		{author} {\bibfnamefont {M.}~\bibnamefont {Aidelsburger}}, \ and\ \bibinfo
		{author} {\bibfnamefont {I.}~\bibnamefont {Bloch}},\ }\href {\doibase
		10.1038/nphys3584} {\bibfield  {journal} {\bibinfo  {journal} {Nat. Phys.}\
		}\textbf {\bibinfo {volume} {12}},\ \bibinfo {pages} {350} (\bibinfo {year}
		{2016})}\BibitemShut {NoStop}%
	\bibitem [{\citenamefont {Wu}\ and\ \citenamefont {Niu}(2003)}]{Wu2003}%
	\BibitemOpen
	\bibfield  {author} {\bibinfo {author} {\bibfnamefont {B.}~\bibnamefont
			{Wu}}\ and\ \bibinfo {author} {\bibfnamefont {Q.}~\bibnamefont {Niu}},\
	}\href {\doibase 10.1088/1367-2630/5/1/104} {\bibfield  {journal} {\bibinfo
			{journal} {New J. Phys.}\ }\textbf {\bibinfo {volume} {5}},\ \bibinfo {pages}
		{104} (\bibinfo {year} {2003})}\BibitemShut {NoStop}%
	\bibitem [{\citenamefont {Tomadin}\ \emph {et~al.}(2008)\citenamefont
		{Tomadin}, \citenamefont {Mannella},\ and\ \citenamefont
		{Wimberger}}]{Tomadin2008}%
	\BibitemOpen
	\bibfield  {author} {\bibinfo {author} {\bibfnamefont {A.}~\bibnamefont
			{Tomadin}}, \bibinfo {author} {\bibfnamefont {R.}~\bibnamefont {Mannella}}, \
		and\ \bibinfo {author} {\bibfnamefont {S.}~\bibnamefont {Wimberger}},\ }\href
	{\doibase 10.1103/PhysRevA.77.013606} {\bibfield  {journal} {\bibinfo
			{journal} {Phys. Rev. A}\ }\textbf {\bibinfo {volume} {77}},\ \bibinfo
		{pages} {013606} (\bibinfo {year} {2008})}\BibitemShut {NoStop}%
	\bibitem [{\citenamefont {Deng}\ \emph {et~al.}(2015)\citenamefont {Deng},
		\citenamefont {Dai}, \citenamefont {Huang}, \citenamefont {Qin},
		\citenamefont {Xu}, \citenamefont {Zhong}, \citenamefont {He},\ and\
		\citenamefont {Lee}}]{Deng2015a}%
	\BibitemOpen
	\bibfield  {author} {\bibinfo {author} {\bibfnamefont {H.}~\bibnamefont
			{Deng}}, \bibinfo {author} {\bibfnamefont {H.}~\bibnamefont {Dai}}, \bibinfo
		{author} {\bibfnamefont {J.}~\bibnamefont {Huang}}, \bibinfo {author}
		{\bibfnamefont {X.}~\bibnamefont {Qin}}, \bibinfo {author} {\bibfnamefont
			{J.}~\bibnamefont {Xu}}, \bibinfo {author} {\bibfnamefont {H.}~\bibnamefont
			{Zhong}}, \bibinfo {author} {\bibfnamefont {C.}~\bibnamefont {He}}, \ and\
		\bibinfo {author} {\bibfnamefont {C.}~\bibnamefont {Lee}},\ }\href {\doibase
		10.1103/PhysRevA.92.023618} {\bibfield  {journal} {\bibinfo  {journal} {Phys.
				Rev. A}\ }\textbf {\bibinfo {volume} {92}},\ \bibinfo {pages} {023618}
		(\bibinfo {year} {2015})}\BibitemShut {NoStop}%
	\bibitem [{\citenamefont {Zurek}\ \emph {et~al.}(2005)\citenamefont {Zurek},
		\citenamefont {Dorner},\ and\ \citenamefont {Zoller}}]{Zurek2005}%
	\BibitemOpen
	\bibfield  {author} {\bibinfo {author} {\bibfnamefont {W.~H.}\ \bibnamefont
			{Zurek}}, \bibinfo {author} {\bibfnamefont {U.}~\bibnamefont {Dorner}}, \
		and\ \bibinfo {author} {\bibfnamefont {P.}~\bibnamefont {Zoller}},\ }\href
	{\doibase 10.1103/PhysRevLett.95.105701} {\bibfield  {journal} {\bibinfo
			{journal} {Phys. Rev. Lett.}\ }\textbf {\bibinfo {volume} {95}},\ \bibinfo
		{pages} {105701} (\bibinfo {year} {2005})}\BibitemShut {NoStop}%
	\bibitem [{\citenamefont {Dziarmaga}\ \emph {et~al.}(2012)\citenamefont
		{Dziarmaga}, \citenamefont {Tylutki},\ and\ \citenamefont
		{Zurek}}]{Dziarmaga2012}%
	\BibitemOpen
	\bibfield  {author} {\bibinfo {author} {\bibfnamefont {J.}~\bibnamefont
			{Dziarmaga}}, \bibinfo {author} {\bibfnamefont {M.}~\bibnamefont {Tylutki}},
		\ and\ \bibinfo {author} {\bibfnamefont {W.~H.}\ \bibnamefont {Zurek}},\
	}\href {\doibase 10.1103/PhysRevB.86.144521} {\bibfield  {journal} {\bibinfo
			{journal} {Phys. Rev. B}\ }\textbf {\bibinfo {volume} {86}},\ \bibinfo
		{pages} {144521} (\bibinfo {year} {2012})}\BibitemShut {NoStop}%
	\bibitem [{\citenamefont {del Campo}\ and\ \citenamefont
		{Zurek}(2014)}]{DelCampo2013}%
	\BibitemOpen
	\bibfield  {author} {\bibinfo {author} {\bibfnamefont {A.}~\bibnamefont {del
				Campo}}\ and\ \bibinfo {author} {\bibfnamefont {W.~H.}\ \bibnamefont
			{Zurek}},\ }\href {\doibase 10.1142/S0217751X1430018X} {\bibfield  {journal}
		{\bibinfo  {journal} {Int. J. Mod. Phys. A}\ }\textbf {\bibinfo {volume}
			{29}},\ \bibinfo {pages} {1430018} (\bibinfo {year} {2014})}\BibitemShut
	{NoStop}%
	\bibitem [{\citenamefont {Shimizu}\ \emph
		{et~al.}(2018{\natexlab{a}})\citenamefont {Shimizu}, \citenamefont {Kuno},
		\citenamefont {Hirano},\ and\ \citenamefont {Ichinose}}]{Shimizu2018a}%
	\BibitemOpen
	\bibfield  {author} {\bibinfo {author} {\bibfnamefont {K.}~\bibnamefont
			{Shimizu}}, \bibinfo {author} {\bibfnamefont {Y.}~\bibnamefont {Kuno}},
		\bibinfo {author} {\bibfnamefont {T.}~\bibnamefont {Hirano}}, \ and\ \bibinfo
		{author} {\bibfnamefont {I.}~\bibnamefont {Ichinose}},\ }\href {\doibase
		10.1103/PhysRevA.97.033626} {\bibfield  {journal} {\bibinfo  {journal} {Phys.
				Rev. A}\ }\textbf {\bibinfo {volume} {97}},\ \bibinfo {pages} {033626}
		(\bibinfo {year} {2018}{\natexlab{a}})}\BibitemShut {NoStop}%
	\bibitem [{\citenamefont {Shimizu}\ \emph
		{et~al.}(2018{\natexlab{b}})\citenamefont {Shimizu}, \citenamefont {Hirano},
		\citenamefont {Park}, \citenamefont {Kuno},\ and\ \citenamefont
		{Ichinose}}]{Shimizu2018b}%
	\BibitemOpen
	\bibfield  {author} {\bibinfo {author} {\bibfnamefont {K.}~\bibnamefont
			{Shimizu}}, \bibinfo {author} {\bibfnamefont {T.}~\bibnamefont {Hirano}},
		\bibinfo {author} {\bibfnamefont {J.}~\bibnamefont {Park}}, \bibinfo {author}
		{\bibfnamefont {Y.}~\bibnamefont {Kuno}}, \ and\ \bibinfo {author}
		{\bibfnamefont {I.}~\bibnamefont {Ichinose}},\ }\href {\doibase
		10.1088/1367-2630/aad5f9} {\bibfield  {journal} {\bibinfo  {journal} {New J.
				Phys.}\ }\textbf {\bibinfo {volume} {20}},\ \bibinfo {pages} {083006}
		(\bibinfo {year} {2018}{\natexlab{b}})}\BibitemShut {NoStop}%
	\bibitem [{\citenamefont {Shimizu}\ \emph
		{et~al.}(2018{\natexlab{c}})\citenamefont {Shimizu}, \citenamefont {Hirano},
		\citenamefont {Park}, \citenamefont {Kuno},\ and\ \citenamefont
		{Ichinose}}]{Shimizu2018c}%
	\BibitemOpen
	\bibfield  {author} {\bibinfo {author} {\bibfnamefont {K.}~\bibnamefont
			{Shimizu}}, \bibinfo {author} {\bibfnamefont {T.}~\bibnamefont {Hirano}},
		\bibinfo {author} {\bibfnamefont {J.}~\bibnamefont {Park}}, \bibinfo {author}
		{\bibfnamefont {Y.}~\bibnamefont {Kuno}}, \ and\ \bibinfo {author}
		{\bibfnamefont {I.}~\bibnamefont {Ichinose}},\ }\href {\doibase
		10.1103/PhysRevA.98.063603} {\bibfield  {journal} {\bibinfo  {journal} {Phys.
				Rev. A}\ }\textbf {\bibinfo {volume} {98}},\ \bibinfo {pages} {063603}
		(\bibinfo {year} {2018}{\natexlab{c}})}\BibitemShut {NoStop}%
	\bibitem [{\citenamefont {Weiss}\ \emph {et~al.}(2018)\citenamefont {Weiss},
		\citenamefont {Gerster}, \citenamefont {Jaschke}, \citenamefont {Silvi},\
		and\ \citenamefont {Montangero}}]{Weiss2018}%
	\BibitemOpen
	\bibfield  {author} {\bibinfo {author} {\bibfnamefont {W.}~\bibnamefont
			{Weiss}}, \bibinfo {author} {\bibfnamefont {M.}~\bibnamefont {Gerster}},
		\bibinfo {author} {\bibfnamefont {D.}~\bibnamefont {Jaschke}}, \bibinfo
		{author} {\bibfnamefont {P.}~\bibnamefont {Silvi}}, \ and\ \bibinfo {author}
		{\bibfnamefont {S.}~\bibnamefont {Montangero}},\ }\href {\doibase
		10.1103/PhysRevA.98.063601} {\bibfield  {journal} {\bibinfo  {journal} {Phys.
				Rev. A}\ }\textbf {\bibinfo {volume} {98}},\ \bibinfo {pages} {063601}
		(\bibinfo {year} {2018})}\BibitemShut {NoStop}%
	\bibitem [{\citenamefont {Scherg}\ \emph {et~al.}(2018)\citenamefont {Scherg},
		\citenamefont {Kohlert}, \citenamefont {Herbrych}, \citenamefont {Stolpp},
		\citenamefont {Bordia}, \citenamefont {Schneider}, \citenamefont
		{Heidrich-Meisner}, \citenamefont {Bloch},\ and\ \citenamefont
		{Aidelsburger}}]{Scherg2018}%
	\BibitemOpen
	\bibfield  {author} {\bibinfo {author} {\bibfnamefont {S.}~\bibnamefont
			{Scherg}}, \bibinfo {author} {\bibfnamefont {T.}~\bibnamefont {Kohlert}},
		\bibinfo {author} {\bibfnamefont {J.}~\bibnamefont {Herbrych}}, \bibinfo
		{author} {\bibfnamefont {J.}~\bibnamefont {Stolpp}}, \bibinfo {author}
		{\bibfnamefont {P.}~\bibnamefont {Bordia}}, \bibinfo {author} {\bibfnamefont
			{U.}~\bibnamefont {Schneider}}, \bibinfo {author} {\bibfnamefont
			{F.}~\bibnamefont {Heidrich-Meisner}}, \bibinfo {author} {\bibfnamefont
			{I.}~\bibnamefont {Bloch}}, \ and\ \bibinfo {author} {\bibfnamefont
			{M.}~\bibnamefont {Aidelsburger}},\ }\href {\doibase
		10.1103/PhysRevLett.121.130402} {\bibfield  {journal} {\bibinfo  {journal}
			{Phys. Rev. Lett.}\ }\textbf {\bibinfo {volume} {121}},\ \bibinfo {pages}
		{130402} (\bibinfo {year} {2018})}\BibitemShut {NoStop}%
	\bibitem [{\citenamefont {Brown}\ \emph {et~al.}(2019)\citenamefont {Brown},
		\citenamefont {Mitra}, \citenamefont {Guardado-Sanchez}, \citenamefont
		{Nourafkan}, \citenamefont {Reymbaut}, \citenamefont {H{\'{e}}bert},
		\citenamefont {Bergeron}, \citenamefont {Tremblay}, \citenamefont {Kokalj},
		\citenamefont {Huse}, \citenamefont {Schau{\ss}},\ and\ \citenamefont
		{Bakr}}]{Brown2018}%
	\BibitemOpen
	\bibfield  {author} {\bibinfo {author} {\bibfnamefont {P.~T.}\ \bibnamefont
			{Brown}}, \bibinfo {author} {\bibfnamefont {D.}~\bibnamefont {Mitra}},
		\bibinfo {author} {\bibfnamefont {E.}~\bibnamefont {Guardado-Sanchez}},
		\bibinfo {author} {\bibfnamefont {R.}~\bibnamefont {Nourafkan}}, \bibinfo
		{author} {\bibfnamefont {A.}~\bibnamefont {Reymbaut}}, \bibinfo {author}
		{\bibfnamefont {C.-D.}\ \bibnamefont {H{\'{e}}bert}}, \bibinfo {author}
		{\bibfnamefont {S.}~\bibnamefont {Bergeron}}, \bibinfo {author}
		{\bibfnamefont {A.-M.~S.}\ \bibnamefont {Tremblay}}, \bibinfo {author}
		{\bibfnamefont {J.}~\bibnamefont {Kokalj}}, \bibinfo {author} {\bibfnamefont
			{D.~A.}\ \bibnamefont {Huse}}, \bibinfo {author} {\bibfnamefont
			{P.}~\bibnamefont {Schau{\ss}}}, \ and\ \bibinfo {author} {\bibfnamefont
			{W.~S.}\ \bibnamefont {Bakr}},\ }\href {\doibase 10.1126/science.aat4134}
	{\bibfield  {journal} {\bibinfo  {journal} {Science}\ }\textbf {\bibinfo
			{volume} {363}},\ \bibinfo {pages} {379} (\bibinfo {year}
		{2019})}\BibitemShut {NoStop}%
	\bibitem [{\citenamefont {Fujiwara}\ \emph {et~al.}(2019)\citenamefont
		{Fujiwara}, \citenamefont {Singh}, \citenamefont {Geiger}, \citenamefont
		{Senaratne}, \citenamefont {Rajagopal}, \citenamefont {Lipatov},\ and\
		\citenamefont {Weld}}]{Fujiwara2019}%
	\BibitemOpen
	\bibfield  {author} {\bibinfo {author} {\bibfnamefont {C.~J.}\ \bibnamefont
			{Fujiwara}}, \bibinfo {author} {\bibfnamefont {K.}~\bibnamefont {Singh}},
		\bibinfo {author} {\bibfnamefont {Z.~A.}\ \bibnamefont {Geiger}}, \bibinfo
		{author} {\bibfnamefont {R.}~\bibnamefont {Senaratne}}, \bibinfo {author}
		{\bibfnamefont {S.~V.}\ \bibnamefont {Rajagopal}}, \bibinfo {author}
		{\bibfnamefont {M.}~\bibnamefont {Lipatov}}, \ and\ \bibinfo {author}
		{\bibfnamefont {D.~M.}\ \bibnamefont {Weld}},\ }\href {\doibase
		10.1103/PhysRevLett.122.010402} {\bibfield  {journal} {\bibinfo  {journal}
			{Phys. Rev. Lett.}\ }\textbf {\bibinfo {volume} {122}},\ \bibinfo {pages}
		{010402} (\bibinfo {year} {2019})}\BibitemShut {NoStop}%
	\bibitem [{\citenamefont {L{\'{e}}onard}\ \emph {et~al.}(2017)\citenamefont
		{L{\'{e}}onard}, \citenamefont {Morales}, \citenamefont {Zupancic},
		\citenamefont {Donner},\ and\ \citenamefont {Esslinger}}]{Leonard2017a}%
	\BibitemOpen
	\bibfield  {author} {\bibinfo {author} {\bibfnamefont {J.}~\bibnamefont
			{L{\'{e}}onard}}, \bibinfo {author} {\bibfnamefont {A.}~\bibnamefont
			{Morales}}, \bibinfo {author} {\bibfnamefont {P.}~\bibnamefont {Zupancic}},
		\bibinfo {author} {\bibfnamefont {T.}~\bibnamefont {Donner}}, \ and\ \bibinfo
		{author} {\bibfnamefont {T.}~\bibnamefont {Esslinger}},\ }\href {\doibase
		10.1126/science.aan2608} {\bibfield  {journal} {\bibinfo  {journal}
			{Science}\ }\textbf {\bibinfo {volume} {358}},\ \bibinfo {pages} {1415}
		(\bibinfo {year} {2017})}\BibitemShut {NoStop}%
	\bibitem [{\citenamefont {{Di Liberto}}\ \emph {et~al.}(2018)\citenamefont {{Di
				Liberto}}, \citenamefont {Recati}, \citenamefont {Trivedi}, \citenamefont
		{Carusotto},\ and\ \citenamefont {Menotti}}]{DiLiberto2018}%
	\BibitemOpen
	\bibfield  {author} {\bibinfo {author} {\bibfnamefont {M.}~\bibnamefont {{Di
					Liberto}}}, \bibinfo {author} {\bibfnamefont {A.}~\bibnamefont {Recati}},
		\bibinfo {author} {\bibfnamefont {N.}~\bibnamefont {Trivedi}}, \bibinfo
		{author} {\bibfnamefont {I.}~\bibnamefont {Carusotto}}, \ and\ \bibinfo
		{author} {\bibfnamefont {C.}~\bibnamefont {Menotti}},\ }\href {\doibase
		10.1103/PhysRevLett.120.073201} {\bibfield  {journal} {\bibinfo  {journal}
			{Phys. Rev. Lett.}\ }\textbf {\bibinfo {volume} {120}},\ \bibinfo {pages}
		{073201} (\bibinfo {year} {2018})}\BibitemShut {NoStop}%
	\bibitem [{\citenamefont {Eckardt}\ \emph {et~al.}(2005)\citenamefont
		{Eckardt}, \citenamefont {Weiss},\ and\ \citenamefont
		{Holthaus}}]{Eckardt2005}%
	\BibitemOpen
	\bibfield  {author} {\bibinfo {author} {\bibfnamefont {A.}~\bibnamefont
			{Eckardt}}, \bibinfo {author} {\bibfnamefont {C.}~\bibnamefont {Weiss}}, \
		and\ \bibinfo {author} {\bibfnamefont {M.}~\bibnamefont {Holthaus}},\ }\href
	{\doibase 10.1103/PhysRevLett.95.260404} {\bibfield  {journal} {\bibinfo
			{journal} {Phys. Rev. Lett.}\ }\textbf {\bibinfo {volume} {95}},\ \bibinfo
		{pages} {260404} (\bibinfo {year} {2005})}\BibitemShut {NoStop}%
	\bibitem [{\citenamefont {Gaul}\ \emph {et~al.}(2009)\citenamefont {Gaul},
		\citenamefont {Lima}, \citenamefont {D{\'{i}}az}, \citenamefont
		{M{\"{u}}ller},\ and\ \citenamefont {Dom{\'{i}}nguez-Adame}}]{Gaul2009}%
	\BibitemOpen
	\bibfield  {author} {\bibinfo {author} {\bibfnamefont {C.}~\bibnamefont
			{Gaul}}, \bibinfo {author} {\bibfnamefont {R.~P.~A.}\ \bibnamefont {Lima}},
		\bibinfo {author} {\bibfnamefont {E.}~\bibnamefont {D{\'{i}}az}}, \bibinfo
		{author} {\bibfnamefont {C.~A.}\ \bibnamefont {M{\"{u}}ller}}, \ and\
		\bibinfo {author} {\bibfnamefont {F.}~\bibnamefont {Dom{\'{i}}nguez-Adame}},\
	}\href {\doibase 10.1103/PhysRevLett.102.255303} {\bibfield  {journal}
		{\bibinfo  {journal} {Phys. Rev. Lett.}\ }\textbf {\bibinfo {volume} {102}},\
		\bibinfo {pages} {255303} (\bibinfo {year} {2009})}\BibitemShut {NoStop}%
	\bibitem [{\citenamefont {Green}\ and\ \citenamefont
		{Sondhi}(2005)}]{Green2005}%
	\BibitemOpen
	\bibfield  {author} {\bibinfo {author} {\bibfnamefont {A.~G.}\ \bibnamefont
			{Green}}\ and\ \bibinfo {author} {\bibfnamefont {S.~L.}\ \bibnamefont
			{Sondhi}},\ }\href {\doibase 10.1103/PhysRevLett.95.267001} {\bibfield
		{journal} {\bibinfo  {journal} {Phys. Rev. Lett.}\ }\textbf {\bibinfo
			{volume} {95}},\ \bibinfo {pages} {267001} (\bibinfo {year}
		{2005})}\BibitemShut {NoStop}%
	\bibitem [{\citenamefont {Polkovnikov}\ \emph {et~al.}(2011)\citenamefont
		{Polkovnikov}, \citenamefont {Sengupta}, \citenamefont {Silva},\ and\
		\citenamefont {Vengalattore}}]{Polkovnikov2011}%
	\BibitemOpen
	\bibfield  {author} {\bibinfo {author} {\bibfnamefont {A.}~\bibnamefont
			{Polkovnikov}}, \bibinfo {author} {\bibfnamefont {K.}~\bibnamefont
			{Sengupta}}, \bibinfo {author} {\bibfnamefont {A.}~\bibnamefont {Silva}}, \
		and\ \bibinfo {author} {\bibfnamefont {M.}~\bibnamefont {Vengalattore}},\
	}\href {\doibase 10.1103/RevModPhys.83.863} {\bibfield  {journal} {\bibinfo
			{journal} {Rev. Mod. Phys.}\ }\textbf {\bibinfo {volume} {83}},\ \bibinfo
		{pages} {863} (\bibinfo {year} {2011})}\BibitemShut {NoStop}%
	\bibitem [{\citenamefont {Kennett}(2013)}]{Kennett2013}%
	\BibitemOpen
	\bibfield  {author} {\bibinfo {author} {\bibfnamefont {M.~P.}\ \bibnamefont
			{Kennett}},\ }\href {\doibase 10.1155/2013/393616} {\bibfield  {journal}
		{\bibinfo  {journal} {ISRN Condens. Matter Phys.}\ }\textbf {\bibinfo
			{volume} {2013}},\ \bibinfo {pages} {1} (\bibinfo {year} {2013})}\BibitemShut
	{NoStop}%
	\bibitem [{\citenamefont {Freericks}\ and\ \citenamefont
		{Monien}(1994)}]{Freericks1994}%
	\BibitemOpen
	\bibfield  {author} {\bibinfo {author} {\bibfnamefont {J.~K.}\ \bibnamefont
			{Freericks}}\ and\ \bibinfo {author} {\bibfnamefont {H.}~\bibnamefont
			{Monien}},\ }\href {\doibase 10.1209/0295-5075/26/7/012} {\bibfield
		{journal} {\bibinfo  {journal} {EPL}\ }\textbf {\bibinfo {volume} {26}},\
		\bibinfo {pages} {545} (\bibinfo {year} {1994})}\BibitemShut {NoStop}%
	\bibitem [{\citenamefont {Elesin}\ \emph {et~al.}(1994)\citenamefont {Elesin},
		\citenamefont {Kashurnikov},\ and\ \citenamefont {Openov}}]{Elesin1994}%
	\BibitemOpen
	\bibfield  {author} {\bibinfo {author} {\bibfnamefont {V.~F.}\ \bibnamefont
			{Elesin}}, \bibinfo {author} {\bibfnamefont {V.~A.}\ \bibnamefont
			{Kashurnikov}}, \ and\ \bibinfo {author} {\bibfnamefont {L.~A.}\ \bibnamefont
			{Openov}},\ }\href {http://jetpletters.ac.ru/ps/1323/article_20004.shtml
		http://www.jetpletters.ac.ru/ps/1323/article_20004.shtml} {\bibfield
		{journal} {\bibinfo  {journal} {JETP Lett.}\ }\textbf {\bibinfo {volume}
			{60}},\ \bibinfo {pages} {177} (\bibinfo {year} {1994})}\BibitemShut
	{NoStop}%
	\bibitem [{\citenamefont {F{\"{o}}lling}\ \emph {et~al.}(2005)\citenamefont
		{F{\"{o}}lling}, \citenamefont {Gerbier}, \citenamefont {Widera},
		\citenamefont {Mandel}, \citenamefont {Gericke},\ and\ \citenamefont
		{Bloch}}]{Folling2005}%
	\BibitemOpen
	\bibfield  {author} {\bibinfo {author} {\bibfnamefont {S.}~\bibnamefont
			{F{\"{o}}lling}}, \bibinfo {author} {\bibfnamefont {F.}~\bibnamefont
			{Gerbier}}, \bibinfo {author} {\bibfnamefont {A.}~\bibnamefont {Widera}},
		\bibinfo {author} {\bibfnamefont {O.}~\bibnamefont {Mandel}}, \bibinfo
		{author} {\bibfnamefont {T.}~\bibnamefont {Gericke}}, \ and\ \bibinfo
		{author} {\bibfnamefont {I.}~\bibnamefont {Bloch}},\ }\href {\doibase
		10.1038/nature03500} {\bibfield  {journal} {\bibinfo  {journal} {Nature}\
		}\textbf {\bibinfo {volume} {434}},\ \bibinfo {pages} {481} (\bibinfo {year}
		{2005})}\BibitemShut {NoStop}%
	\bibitem [{\citenamefont {K{\"{u}}hner}\ \emph {et~al.}(2000)\citenamefont
		{K{\"{u}}hner}, \citenamefont {White},\ and\ \citenamefont
		{Monien}}]{Kuhner2000}%
	\BibitemOpen
	\bibfield  {author} {\bibinfo {author} {\bibfnamefont {T.~D.}\ \bibnamefont
			{K{\"{u}}hner}}, \bibinfo {author} {\bibfnamefont {S.~R.}\ \bibnamefont
			{White}}, \ and\ \bibinfo {author} {\bibfnamefont {H.}~\bibnamefont
			{Monien}},\ }\href {\doibase 10.1103/PhysRevB.61.12474} {\bibfield  {journal}
		{\bibinfo  {journal} {Phys. Rev. B}\ }\textbf {\bibinfo {volume} {61}},\
		\bibinfo {pages} {12474} (\bibinfo {year} {2000})}\BibitemShut {NoStop}%
	\bibitem [{\citenamefont {Rossini}\ and\ \citenamefont
		{Fazio}(2012)}]{Rossini2012}%
	\BibitemOpen
	\bibfield  {author} {\bibinfo {author} {\bibfnamefont {D.}~\bibnamefont
			{Rossini}}\ and\ \bibinfo {author} {\bibfnamefont {R.}~\bibnamefont
			{Fazio}},\ }\href {\doibase 10.1088/1367-2630/14/6/065012} {\bibfield
		{journal} {\bibinfo  {journal} {New J. Phys.}\ }\textbf {\bibinfo {volume}
			{14}},\ \bibinfo {pages} {065012} (\bibinfo {year} {2012})}\BibitemShut
	{NoStop}%
	\bibitem [{\citenamefont {G{\'{o}}ral}\ \emph {et~al.}(2000)\citenamefont
		{G{\'{o}}ral}, \citenamefont {Rza{\k{a}}{\.{z}}ewski},\ and\ \citenamefont
		{Pfau}}]{Goral2000}%
	\BibitemOpen
	\bibfield  {author} {\bibinfo {author} {\bibfnamefont {K.}~\bibnamefont
			{G{\'{o}}ral}}, \bibinfo {author} {\bibfnamefont {K.}~\bibnamefont
			{Rz{\k{a}}{\.{z}}ewski}}, \ and\ \bibinfo {author} {\bibfnamefont
			{T.}~\bibnamefont {Pfau}},\ }\href {\doibase 10.1103/PhysRevA.61.051601}
	{\bibfield  {journal} {\bibinfo  {journal} {Phys. Rev. A}\ }\textbf {\bibinfo
			{volume} {61}},\ \bibinfo {pages} {051601(R)} (\bibinfo {year}
		{2000})}\BibitemShut {NoStop}%
	\bibitem [{\citenamefont {Griesmaier}\ \emph {et~al.}(2005)\citenamefont
		{Griesmaier}, \citenamefont {Werner}, \citenamefont {Hensler}, \citenamefont
		{Stuhler},\ and\ \citenamefont {Pfau}}]{Griesmaier2005}%
	\BibitemOpen
	\bibfield  {author} {\bibinfo {author} {\bibfnamefont {A.}~\bibnamefont
			{Griesmaier}}, \bibinfo {author} {\bibfnamefont {J.}~\bibnamefont {Werner}},
		\bibinfo {author} {\bibfnamefont {S.}~\bibnamefont {Hensler}}, \bibinfo
		{author} {\bibfnamefont {J.}~\bibnamefont {Stuhler}}, \ and\ \bibinfo
		{author} {\bibfnamefont {T.}~\bibnamefont {Pfau}},\ }\href {\doibase
		10.1103/PhysRevLett.94.160401} {\bibfield  {journal} {\bibinfo  {journal}
			{Phys. Rev. Lett.}\ }\textbf {\bibinfo {volume} {94}},\ \bibinfo {pages}
		{160401} (\bibinfo {year} {2005})}\BibitemShut {NoStop}%
	\bibitem [{\citenamefont {Lu}\ \emph {et~al.}(2011)\citenamefont {Lu},
		\citenamefont {Burdick}, \citenamefont {Youn},\ and\ \citenamefont
		{Lev}}]{Lu2011}%
	\BibitemOpen
	\bibfield  {author} {\bibinfo {author} {\bibfnamefont {M.}~\bibnamefont
			{Lu}}, \bibinfo {author} {\bibfnamefont {N.~Q.}\ \bibnamefont {Burdick}},
		\bibinfo {author} {\bibfnamefont {S.~H.}\ \bibnamefont {Youn}}, \ and\
		\bibinfo {author} {\bibfnamefont {B.~L.}\ \bibnamefont {Lev}},\ }\href
	{\doibase 10.1103/PhysRevLett.107.190401} {\bibfield  {journal} {\bibinfo
			{journal} {Phys. Rev. Lett.}\ }\textbf {\bibinfo {volume} {107}},\ \bibinfo
		{pages} {190401} (\bibinfo {year} {2011})}\BibitemShut {NoStop}%
	\bibitem [{\citenamefont {Aikawa}\ \emph {et~al.}(2012)\citenamefont {Aikawa},
		\citenamefont {Frisch}, \citenamefont {Mark}, \citenamefont {Baier},
		\citenamefont {Rietzler}, \citenamefont {Grimm},\ and\ \citenamefont
		{Ferlaino}}]{Aikawa2012}%
	\BibitemOpen
	\bibfield  {author} {\bibinfo {author} {\bibfnamefont {K.}~\bibnamefont
			{Aikawa}}, \bibinfo {author} {\bibfnamefont {A.}~\bibnamefont {Frisch}},
		\bibinfo {author} {\bibfnamefont {M.}~\bibnamefont {Mark}}, \bibinfo {author}
		{\bibfnamefont {S.}~\bibnamefont {Baier}}, \bibinfo {author} {\bibfnamefont
			{A.}~\bibnamefont {Rietzler}}, \bibinfo {author} {\bibfnamefont
			{R.}~\bibnamefont {Grimm}}, \ and\ \bibinfo {author} {\bibfnamefont
			{F.}~\bibnamefont {Ferlaino}},\ }\href {\doibase
		10.1103/PhysRevLett.108.210401} {\bibfield  {journal} {\bibinfo  {journal}
			{Phys. Rev. Lett.}\ }\textbf {\bibinfo {volume} {108}},\ \bibinfo {pages}
		{210401} (\bibinfo {year} {2012})}\BibitemShut {NoStop}%
	\bibitem [{\citenamefont {B{\"{u}}chler}\ \emph {et~al.}(2007)\citenamefont
		{B{\"{u}}chler}, \citenamefont {Micheli},\ and\ \citenamefont
		{Zoller}}]{Buchler2007}%
	\BibitemOpen
	\bibfield  {author} {\bibinfo {author} {\bibfnamefont {H.~P.}\ \bibnamefont
			{B{\"{u}}chler}}, \bibinfo {author} {\bibfnamefont {A.}~\bibnamefont
			{Micheli}}, \ and\ \bibinfo {author} {\bibfnamefont {P.}~\bibnamefont
			{Zoller}},\ }\href {\doibase 10.1038/nphys678} {\bibfield  {journal}
		{\bibinfo  {journal} {Nat. Phys.}\ }\textbf {\bibinfo {volume} {3}},\
		\bibinfo {pages} {726} (\bibinfo {year} {2007})}\BibitemShut {NoStop}%
	\bibitem [{\citenamefont {Micheli}\ \emph {et~al.}(2005)\citenamefont
		{Micheli}, \citenamefont {Brennen},\ and\ \citenamefont
		{Zoller}}]{Micheli2006}%
	\BibitemOpen
	\bibfield  {author} {\bibinfo {author} {\bibfnamefont {A.}~\bibnamefont
			{Micheli}}, \bibinfo {author} {\bibfnamefont {G.~K.}\ \bibnamefont
			{Brennen}}, \ and\ \bibinfo {author} {\bibfnamefont {P.}~\bibnamefont
			{Zoller}},\ }\href {\doibase 10.1038/nphys287} {\bibfield  {journal}
		{\bibinfo  {journal} {Nat. Phys.}\ }\textbf {\bibinfo {volume} {2}},\
		\bibinfo {pages} {341} (\bibinfo {year} {2005})}\BibitemShut {NoStop}%
	\bibitem [{\citenamefont {Kotochigova}\ and\ \citenamefont
		{Tiesinga}(2006)}]{Kotochigova2006}%
	\BibitemOpen
	\bibfield  {author} {\bibinfo {author} {\bibfnamefont {S.}~\bibnamefont
			{Kotochigova}}\ and\ \bibinfo {author} {\bibfnamefont {E.}~\bibnamefont
			{Tiesinga}},\ }\href {\doibase 10.1103/PhysRevA.73.041405} {\bibfield
		{journal} {\bibinfo  {journal} {Phys. Rev. A}\ }\textbf {\bibinfo {volume}
			{73}},\ \bibinfo {pages} {041405(R)} (\bibinfo {year} {2006})}\BibitemShut
	{NoStop}%
	\bibitem [{\citenamefont {Gorshkov}\ \emph {et~al.}(2011)\citenamefont
		{Gorshkov}, \citenamefont {Manmana}, \citenamefont {Chen}, \citenamefont
		{Ye}, \citenamefont {Demler}, \citenamefont {Lukin},\ and\ \citenamefont
		{Rey}}]{Gorshkov2011}%
	\BibitemOpen
	\bibfield  {author} {\bibinfo {author} {\bibfnamefont {A.~V.}\ \bibnamefont
			{Gorshkov}}, \bibinfo {author} {\bibfnamefont {S.~R.}\ \bibnamefont
			{Manmana}}, \bibinfo {author} {\bibfnamefont {G.}~\bibnamefont {Chen}},
		\bibinfo {author} {\bibfnamefont {J.}~\bibnamefont {Ye}}, \bibinfo {author}
		{\bibfnamefont {E.}~\bibnamefont {Demler}}, \bibinfo {author} {\bibfnamefont
			{M.~D.}\ \bibnamefont {Lukin}}, \ and\ \bibinfo {author} {\bibfnamefont
			{A.~M.}\ \bibnamefont {Rey}},\ }\href {\doibase
		10.1103/PhysRevLett.107.115301} {\bibfield  {journal} {\bibinfo  {journal}
			{Phys. Rev. Lett.}\ }\textbf {\bibinfo {volume} {107}},\ \bibinfo {pages}
		{115301} (\bibinfo {year} {2011})}\BibitemShut {NoStop}%
	\bibitem [{\citenamefont {Baier}\ \emph {et~al.}(2016)\citenamefont {Baier},
		\citenamefont {Mark}, \citenamefont {Petter}, \citenamefont {Aikawa},
		\citenamefont {Chomaz}, \citenamefont {Cai}, \citenamefont {Baranov},
		\citenamefont {Zoller},\ and\ \citenamefont {Ferlaino}}]{Baier2016}%
	\BibitemOpen
	\bibfield  {author} {\bibinfo {author} {\bibfnamefont {S.}~\bibnamefont
			{Baier}}, \bibinfo {author} {\bibfnamefont {M.~J.}\ \bibnamefont {Mark}},
		\bibinfo {author} {\bibfnamefont {D.}~\bibnamefont {Petter}}, \bibinfo
		{author} {\bibfnamefont {K.}~\bibnamefont {Aikawa}}, \bibinfo {author}
		{\bibfnamefont {L.}~\bibnamefont {Chomaz}}, \bibinfo {author} {\bibfnamefont
			{Z.}~\bibnamefont {Cai}}, \bibinfo {author} {\bibfnamefont {M.}~\bibnamefont
			{Baranov}}, \bibinfo {author} {\bibfnamefont {P.}~\bibnamefont {Zoller}}, \
		and\ \bibinfo {author} {\bibfnamefont {F.}~\bibnamefont {Ferlaino}},\ }\href
	{\doibase 10.1126/science.aac9812} {\bibfield  {journal} {\bibinfo  {journal}
			{Science}\ }\textbf {\bibinfo {volume} {352}},\ \bibinfo {pages} {201}
		(\bibinfo {year} {2016})}\BibitemShut {NoStop}%
	\bibitem [{\citenamefont {Balewski}\ \emph {et~al.}(2014)\citenamefont
		{Balewski}, \citenamefont {Krupp}, \citenamefont {Gaj}, \citenamefont
		{Hofferberth}, \citenamefont {L{\"{o}}w},\ and\ \citenamefont
		{Pfau}}]{Balewski2014}%
	\BibitemOpen
	\bibfield  {author} {\bibinfo {author} {\bibfnamefont {J.~B.}\ \bibnamefont
			{Balewski}}, \bibinfo {author} {\bibfnamefont {A.~T.}\ \bibnamefont {Krupp}},
		\bibinfo {author} {\bibfnamefont {A.}~\bibnamefont {Gaj}}, \bibinfo {author}
		{\bibfnamefont {S.}~\bibnamefont {Hofferberth}}, \bibinfo {author}
		{\bibfnamefont {R.}~\bibnamefont {L{\"{o}}w}}, \ and\ \bibinfo {author}
		{\bibfnamefont {T.}~\bibnamefont {Pfau}},\ }\href {\doibase
		10.1088/1367-2630/16/6/063012} {\bibfield  {journal} {\bibinfo  {journal}
			{New J. Phys.}\ }\textbf {\bibinfo {volume} {16}},\ \bibinfo {pages} {063012}
		(\bibinfo {year} {2014})}\BibitemShut {NoStop}%
	\bibitem [{\citenamefont {Jau}\ \emph {et~al.}(2016)\citenamefont {Jau},
		\citenamefont {Hankin}, \citenamefont {Keating}, \citenamefont {Deutsch},\
		and\ \citenamefont {Biedermann}}]{Jau2016}%
	\BibitemOpen
	\bibfield  {author} {\bibinfo {author} {\bibfnamefont {Y.-Y.}\ \bibnamefont
			{Jau}}, \bibinfo {author} {\bibfnamefont {A.~M.}\ \bibnamefont {Hankin}},
		\bibinfo {author} {\bibfnamefont {T.}~\bibnamefont {Keating}}, \bibinfo
		{author} {\bibfnamefont {I.~H.}\ \bibnamefont {Deutsch}}, \ and\ \bibinfo
		{author} {\bibfnamefont {G.~W.}\ \bibnamefont {Biedermann}},\ }\href
	{\doibase 10.1038/nphys3487} {\bibfield  {journal} {\bibinfo  {journal} {Nat.
				Phys.}\ }\textbf {\bibinfo {volume} {12}},\ \bibinfo {pages} {71} (\bibinfo
		{year} {2016})}\BibitemShut {NoStop}%
	\bibitem [{\citenamefont {Zeiher}\ \emph {et~al.}(2016)\citenamefont {Zeiher},
		\citenamefont {van Bijnen}, \citenamefont {Schau{\ss}}, \citenamefont {Hild},
		\citenamefont {Choi}, \citenamefont {Pohl}, \citenamefont {Bloch},\ and\
		\citenamefont {Gross}}]{Zeiher2016}%
	\BibitemOpen
	\bibfield  {author} {\bibinfo {author} {\bibfnamefont {J.}~\bibnamefont
			{Zeiher}}, \bibinfo {author} {\bibfnamefont {R.}~\bibnamefont {van Bijnen}},
		\bibinfo {author} {\bibfnamefont {P.}~\bibnamefont {Schau{\ss}}}, \bibinfo
		{author} {\bibfnamefont {S.}~\bibnamefont {Hild}}, \bibinfo {author}
		{\bibfnamefont {J.-y.}\ \bibnamefont {Choi}}, \bibinfo {author}
		{\bibfnamefont {T.}~\bibnamefont {Pohl}}, \bibinfo {author} {\bibfnamefont
			{I.}~\bibnamefont {Bloch}}, \ and\ \bibinfo {author} {\bibfnamefont
			{C.}~\bibnamefont {Gross}},\ }\href {\doibase 10.1038/nphys3835} {\bibfield
		{journal} {\bibinfo  {journal} {Nat. Phys.}\ }\textbf {\bibinfo {volume}
			{12}},\ \bibinfo {pages} {1095} (\bibinfo {year} {2016})}\BibitemShut
	{NoStop}%
	\bibitem [{\citenamefont {Mukherjee}\ \emph {et~al.}(2016)\citenamefont
		{Mukherjee}, \citenamefont {Killian},\ and\ \citenamefont
		{Hazzard}}]{Mukherjee2016}%
	\BibitemOpen
	\bibfield  {author} {\bibinfo {author} {\bibfnamefont {R.}~\bibnamefont
			{Mukherjee}}, \bibinfo {author} {\bibfnamefont {T.~C.}\ \bibnamefont
			{Killian}}, \ and\ \bibinfo {author} {\bibfnamefont {K.~R.~A.}\ \bibnamefont
			{Hazzard}},\ }\href {\doibase 10.1103/PhysRevA.94.053422} {\bibfield
		{journal} {\bibinfo  {journal} {Phys. Rev. A}\ }\textbf {\bibinfo {volume}
			{94}},\ \bibinfo {pages} {053422} (\bibinfo {year} {2016})}\BibitemShut
	{NoStop}%
	\bibitem [{\citenamefont {Ko{\'{s}}cik}\ and\ \citenamefont
		{Sowi{\'{n}}ski}(2018)}]{Koscik2018}%
	\BibitemOpen
	\bibfield  {author} {\bibinfo {author} {\bibfnamefont {P.}~\bibnamefont
			{Ko{\'{s}}cik}}\ and\ \bibinfo {author} {\bibfnamefont {T.}~\bibnamefont
			{Sowi{\'{n}}ski}},\ }\href {\doibase 10.1038/s41598-017-18505-5} {\bibfield
		{journal} {\bibinfo  {journal} {Sci. Rep.}\ }\textbf {\bibinfo {volume}
			{8}},\ \bibinfo {pages} {48} (\bibinfo {year} {2018})}\BibitemShut {NoStop}%
	\bibitem [{\citenamefont {Ko{\'{s}}cik}\ and\ \citenamefont
		{Sowi{\'{n}}ski}(2019)}]{Koscik2019}%
	\BibitemOpen
	\bibfield  {author} {\bibinfo {author} {\bibfnamefont {P.}~\bibnamefont
			{Ko{\'{s}}cik}}\ and\ \bibinfo {author} {\bibfnamefont {T.}~\bibnamefont
			{Sowi{\'{n}}ski}},\ }\href {\doibase 10.1038/s41598-019-48442-4} {\bibfield
		{journal} {\bibinfo  {journal} {Sci. Rep.}\ }\textbf {\bibinfo {volume}
			{9}},\ \bibinfo {pages} {12018} (\bibinfo {year} {2019})}\BibitemShut
	{NoStop}%
	\bibitem [{\citenamefont {Henkel}\ \emph {et~al.}(2010)\citenamefont {Henkel},
		\citenamefont {Nath},\ and\ \citenamefont {Pohl}}]{Henkel2010}%
	\BibitemOpen
	\bibfield  {author} {\bibinfo {author} {\bibfnamefont {N.}~\bibnamefont
			{Henkel}}, \bibinfo {author} {\bibfnamefont {R.}~\bibnamefont {Nath}}, \ and\
		\bibinfo {author} {\bibfnamefont {T.}~\bibnamefont {Pohl}},\ }\href {\doibase
		10.1103/PhysRevLett.104.195302} {\bibfield  {journal} {\bibinfo  {journal}
			{Phys. Rev. Lett.}\ }\textbf {\bibinfo {volume} {104}},\ \bibinfo {pages}
		{195302} (\bibinfo {year} {2010})}\BibitemShut {NoStop}%
	\bibitem [{\citenamefont {Honer}\ \emph {et~al.}(2010)\citenamefont {Honer},
		\citenamefont {Weimer}, \citenamefont {Pfau},\ and\ \citenamefont
		{B{\"{u}}chler}}]{Honer2010}%
	\BibitemOpen
	\bibfield  {author} {\bibinfo {author} {\bibfnamefont {J.}~\bibnamefont
			{Honer}}, \bibinfo {author} {\bibfnamefont {H.}~\bibnamefont {Weimer}},
		\bibinfo {author} {\bibfnamefont {T.}~\bibnamefont {Pfau}}, \ and\ \bibinfo
		{author} {\bibfnamefont {H.~P.}\ \bibnamefont {B{\"{u}}chler}},\ }\href
	{\doibase 10.1103/PhysRevLett.105.160404} {\bibfield  {journal} {\bibinfo
			{journal} {Phys. Rev. Lett.}\ }\textbf {\bibinfo {volume} {105}},\ \bibinfo
		{pages} {160404} (\bibinfo {year} {2010})}\BibitemShut {NoStop}%
	\bibitem [{\citenamefont {Pupillo}\ \emph {et~al.}(2010)\citenamefont
		{Pupillo}, \citenamefont {Micheli}, \citenamefont {Boninsegni}, \citenamefont
		{Lesanovsky},\ and\ \citenamefont {Zoller}}]{Pupillo2010}%
	\BibitemOpen
	\bibfield  {author} {\bibinfo {author} {\bibfnamefont {G.}~\bibnamefont
			{Pupillo}}, \bibinfo {author} {\bibfnamefont {A.}~\bibnamefont {Micheli}},
		\bibinfo {author} {\bibfnamefont {M.}~\bibnamefont {Boninsegni}}, \bibinfo
		{author} {\bibfnamefont {I.}~\bibnamefont {Lesanovsky}}, \ and\ \bibinfo
		{author} {\bibfnamefont {P.}~\bibnamefont {Zoller}},\ }\href {\doibase
		10.1103/PhysRevLett.104.223002} {\bibfield  {journal} {\bibinfo  {journal}
			{Phys. Rev. Lett.}\ }\textbf {\bibinfo {volume} {104}},\ \bibinfo {pages}
		{223002} (\bibinfo {year} {2010})}\BibitemShut {NoStop}%
	\bibitem [{\citenamefont {Chougale}\ and\ \citenamefont
		{Nath}(2016)}]{Chougale2016}%
	\BibitemOpen
	\bibfield  {author} {\bibinfo {author} {\bibfnamefont {Y.}~\bibnamefont
			{Chougale}}\ and\ \bibinfo {author} {\bibfnamefont {R.}~\bibnamefont
			{Nath}},\ }\href {\doibase 10.1088/0953-4075/49/14/144005} {\bibfield
		{journal} {\bibinfo  {journal} {J. Phys. B}\ }\textbf {\bibinfo {volume}
			{49}},\ \bibinfo {pages} {144005} (\bibinfo {year} {2016})}\BibitemShut
	{NoStop}%
	\bibitem [{\citenamefont {Greiner}\ \emph
		{et~al.}(2002{\natexlab{b}})\citenamefont {Greiner}, \citenamefont {Mandel},
		\citenamefont {H{\"{a}}nsch},\ and\ \citenamefont {Bloch}}]{Greiner2002b}%
	\BibitemOpen
	\bibfield  {author} {\bibinfo {author} {\bibfnamefont {M.}~\bibnamefont
			{Greiner}}, \bibinfo {author} {\bibfnamefont {O.}~\bibnamefont {Mandel}},
		\bibinfo {author} {\bibfnamefont {T.~W.}\ \bibnamefont {H{\"{a}}nsch}}, \
		and\ \bibinfo {author} {\bibfnamefont {I.}~\bibnamefont {Bloch}},\ }\href
	{\doibase 10.1038/nature00968} {\bibfield  {journal} {\bibinfo  {journal}
			{Nature}\ }\textbf {\bibinfo {volume} {419}},\ \bibinfo {pages} {51}
		(\bibinfo {year} {2002}{\natexlab{b}})}\BibitemShut {NoStop}%
	\bibitem [{\citenamefont {Altman}\ \emph {et~al.}(2004)\citenamefont {Altman},
		\citenamefont {Demler},\ and\ \citenamefont {Lukin}}]{Altman2004}%
	\BibitemOpen
	\bibfield  {author} {\bibinfo {author} {\bibfnamefont {E.}~\bibnamefont
			{Altman}}, \bibinfo {author} {\bibfnamefont {E.}~\bibnamefont {Demler}}, \
		and\ \bibinfo {author} {\bibfnamefont {M.~D.}\ \bibnamefont {Lukin}},\ }\href
	{\doibase 10.1103/PhysRevA.70.013603} {\bibfield  {journal} {\bibinfo
			{journal} {Phys. Rev. A}\ }\textbf {\bibinfo {volume} {70}},\ \bibinfo
		{pages} {013603} (\bibinfo {year} {2004})}\BibitemShut {NoStop}%
	\bibitem [{\citenamefont {Rom}\ \emph {et~al.}(2006)\citenamefont {Rom},
		\citenamefont {Best}, \citenamefont {{Van Oosten}}, \citenamefont
		{Schneider}, \citenamefont {F{\"{o}}lling}, \citenamefont {Paredes},\ and\
		\citenamefont {Bloch}}]{Rom2006}%
	\BibitemOpen
	\bibfield  {author} {\bibinfo {author} {\bibfnamefont {T.}~\bibnamefont
			{Rom}}, \bibinfo {author} {\bibfnamefont {T.}~\bibnamefont {Best}}, \bibinfo
		{author} {\bibfnamefont {D.}~\bibnamefont {{Van Oosten}}}, \bibinfo {author}
		{\bibfnamefont {U.}~\bibnamefont {Schneider}}, \bibinfo {author}
		{\bibfnamefont {S.}~\bibnamefont {F{\"{o}}lling}}, \bibinfo {author}
		{\bibfnamefont {B.}~\bibnamefont {Paredes}}, \ and\ \bibinfo {author}
		{\bibfnamefont {I.}~\bibnamefont {Bloch}},\ }\href {\doibase
		10.1038/nature05319} {\bibfield  {journal} {\bibinfo  {journal} {Nature}\
		}\textbf {\bibinfo {volume} {444}},\ \bibinfo {pages} {733} (\bibinfo {year}
		{2006})}\BibitemShut {NoStop}%
	\bibitem [{\citenamefont {Jeltes}\ \emph {et~al.}(2007)\citenamefont {Jeltes},
		\citenamefont {McNamara}, \citenamefont {Hogervorst}, \citenamefont {Vassen},
		\citenamefont {Krachmalnicoff}, \citenamefont {Schellekens}, \citenamefont
		{Perrin}, \citenamefont {Chang}, \citenamefont {Boiron}, \citenamefont
		{Aspect},\ and\ \citenamefont {Westbrook}}]{Jeltes2007}%
	\BibitemOpen
	\bibfield  {author} {\bibinfo {author} {\bibfnamefont {T.}~\bibnamefont
			{Jeltes}}, \bibinfo {author} {\bibfnamefont {J.~M.}\ \bibnamefont
			{McNamara}}, \bibinfo {author} {\bibfnamefont {W.}~\bibnamefont
			{Hogervorst}}, \bibinfo {author} {\bibfnamefont {W.}~\bibnamefont {Vassen}},
		\bibinfo {author} {\bibfnamefont {V.}~\bibnamefont {Krachmalnicoff}},
		\bibinfo {author} {\bibfnamefont {M.}~\bibnamefont {Schellekens}}, \bibinfo
		{author} {\bibfnamefont {A.}~\bibnamefont {Perrin}}, \bibinfo {author}
		{\bibfnamefont {H.}~\bibnamefont {Chang}}, \bibinfo {author} {\bibfnamefont
			{D.}~\bibnamefont {Boiron}}, \bibinfo {author} {\bibfnamefont
			{A.}~\bibnamefont {Aspect}}, \ and\ \bibinfo {author} {\bibfnamefont {C.~I.}\
			\bibnamefont {Westbrook}},\ }\href {\doibase 10.1038/nature05513} {\bibfield
		{journal} {\bibinfo  {journal} {Nature}\ }\textbf {\bibinfo {volume} {445}},\
		\bibinfo {pages} {402} (\bibinfo {year} {2007})}\BibitemShut {NoStop}%
	\bibitem [{\citenamefont {Toth}\ \emph {et~al.}(2008)\citenamefont {Toth},
		\citenamefont {Rey},\ and\ \citenamefont {Blakie}}]{Toth2008}%
	\BibitemOpen
	\bibfield  {author} {\bibinfo {author} {\bibfnamefont {E.}~\bibnamefont
			{Toth}}, \bibinfo {author} {\bibfnamefont {A.~M.}\ \bibnamefont {Rey}}, \
		and\ \bibinfo {author} {\bibfnamefont {P.~B.}\ \bibnamefont {Blakie}},\
	}\href {\doibase 10.1103/PhysRevA.78.013627} {\bibfield  {journal} {\bibinfo
			{journal} {Phys. Rev. A}\ }\textbf {\bibinfo {volume} {78}},\ \bibinfo
		{pages} {013627} (\bibinfo {year} {2008})}\BibitemShut {NoStop}%
	\bibitem [{\citenamefont {Hu}\ \emph {et~al.}(2010)\citenamefont {Hu},
		\citenamefont {Mathey}, \citenamefont {Williams},\ and\ \citenamefont
		{Clark}}]{Hu2010}%
	\BibitemOpen
	\bibfield  {author} {\bibinfo {author} {\bibfnamefont {A.}~\bibnamefont
			{Hu}}, \bibinfo {author} {\bibfnamefont {L.}~\bibnamefont {Mathey}}, \bibinfo
		{author} {\bibfnamefont {C.~J.}\ \bibnamefont {Williams}}, \ and\ \bibinfo
		{author} {\bibfnamefont {C.~W.}\ \bibnamefont {Clark}},\ }\href {\doibase
		10.1103/PhysRevA.81.063602} {\bibfield  {journal} {\bibinfo  {journal} {Phys.
				Rev. A}\ }\textbf {\bibinfo {volume} {81}},\ \bibinfo {pages} {063602}
		(\bibinfo {year} {2010})}\BibitemShut {NoStop}%
	\bibitem [{\citenamefont {Krauth}\ \emph {et~al.}(1992)\citenamefont {Krauth},
		\citenamefont {Caffarel},\ and\ \citenamefont {Bouchaud}}]{Krauth1992}%
	\BibitemOpen
	\bibfield  {author} {\bibinfo {author} {\bibfnamefont {W.}~\bibnamefont
			{Krauth}}, \bibinfo {author} {\bibfnamefont {M.}~\bibnamefont {Caffarel}}, \
		and\ \bibinfo {author} {\bibfnamefont {J.-P.}\ \bibnamefont {Bouchaud}},\
	}\href {\doibase 10.1103/PhysRevB.45.3137} {\bibfield  {journal} {\bibinfo
			{journal} {Phys. Rev. B}\ }\textbf {\bibinfo {volume} {45}},\ \bibinfo
		{pages} {3137} (\bibinfo {year} {1992})}\BibitemShut {NoStop}%
	\bibitem [{\citenamefont {Seibold}\ and\ \citenamefont
		{Lorenzana}(2001)}]{Seibold2001}%
	\BibitemOpen
	\bibfield  {author} {\bibinfo {author} {\bibfnamefont {G.}~\bibnamefont
			{Seibold}}\ and\ \bibinfo {author} {\bibfnamefont {J.}~\bibnamefont
			{Lorenzana}},\ }\href {\doibase 10.1103/PhysRevLett.86.2605} {\bibfield
		{journal} {\bibinfo  {journal} {Phys. Rev. Lett.}\ }\textbf {\bibinfo
			{volume} {86}},\ \bibinfo {pages} {2605} (\bibinfo {year}
		{2001})}\BibitemShut {NoStop}%
	\bibitem [{\citenamefont {Schir{\'{o}}}\ and\ \citenamefont
		{Fabrizio}(2010)}]{Schiro2010}%
	\BibitemOpen
	\bibfield  {author} {\bibinfo {author} {\bibfnamefont {M.}~\bibnamefont
			{Schir{\'{o}}}}\ and\ \bibinfo {author} {\bibfnamefont {M.}~\bibnamefont
			{Fabrizio}},\ }\href {\doibase 10.1103/PhysRevLett.105.076401} {\bibfield
		{journal} {\bibinfo  {journal} {Phys. Rev. Lett.}\ }\textbf {\bibinfo
			{volume} {105}},\ \bibinfo {pages} {076401} (\bibinfo {year}
		{2010})}\BibitemShut {NoStop}%
	\bibitem [{\citenamefont {von Oelsen}\ \emph {et~al.}(2011)\citenamefont {von
			Oelsen}, \citenamefont {Seibold},\ and\ \citenamefont
		{B{\"{u}}nemann}}]{VonOelsen2011a}%
	\BibitemOpen
	\bibfield  {author} {\bibinfo {author} {\bibfnamefont {E.}~\bibnamefont {von
				Oelsen}}, \bibinfo {author} {\bibfnamefont {G.}~\bibnamefont {Seibold}}, \
		and\ \bibinfo {author} {\bibfnamefont {J.}~\bibnamefont {B{\"{u}}nemann}},\
	}\href {\doibase 10.1088/1367-2630/13/11/113031} {\bibfield  {journal}
		{\bibinfo  {journal} {New J. Phys.}\ }\textbf {\bibinfo {volume} {13}},\
		\bibinfo {pages} {113031} (\bibinfo {year} {2011})}\BibitemShut {NoStop}%
\bibitem [{\citenamefont {Bandyopadhyay}\ \emph {et~al.}(2019)\citenamefont
	{Bandyopadhyay}, \citenamefont {Bai}, \citenamefont {Pal}, \citenamefont
	{Suthar}, \citenamefont {Nath},\ and\ \citenamefont
	{Angom}}]{bandyopadhyay_quantum_2019}%
\BibitemOpen
\bibfield  {author} {\bibinfo {author} {\bibfnamefont {S.}~\bibnamefont
		{Bandyopadhyay}}, \bibinfo {author} {\bibfnamefont {R.}~\bibnamefont {Bai}},
	\bibinfo {author} {\bibfnamefont {S.}~\bibnamefont {Pal}}, \bibinfo {author}
	{\bibfnamefont {K.}~\bibnamefont {Suthar}}, \bibinfo {author} {\bibfnamefont
		{R.}~\bibnamefont {Nath}}, \ and\ \bibinfo {author} {\bibfnamefont
		{D.}~\bibnamefont {Angom}},\ }\href {\doibase 10.1103/PhysRevA.100.053623}
{\bibfield  {journal} {\bibinfo  {journal} {Phys. Rev. A}\ }\textbf {\bibinfo
		{volume} {100}},\ \bibinfo {pages} {053623} (\bibinfo {year}
	{2019})}\BibitemShut {NoStop}%
	\bibitem [{\citenamefont {Rey}\ \emph {et~al.}(2003)\citenamefont {Rey},
		\citenamefont {Burnett}, \citenamefont {Roth}, \citenamefont {Edwards},
		\citenamefont {Williams},\ and\ \citenamefont {Clark}}]{Rey2003}%
	\BibitemOpen
	\bibfield  {author} {\bibinfo {author} {\bibfnamefont {A.~M.}\ \bibnamefont
			{Rey}}, \bibinfo {author} {\bibfnamefont {K.}~\bibnamefont {Burnett}},
		\bibinfo {author} {\bibfnamefont {R.}~\bibnamefont {Roth}}, \bibinfo {author}
		{\bibfnamefont {M.}~\bibnamefont {Edwards}}, \bibinfo {author} {\bibfnamefont
			{C.~J.}\ \bibnamefont {Williams}}, \ and\ \bibinfo {author} {\bibfnamefont
			{C.~W.}\ \bibnamefont {Clark}},\ }\href {\doibase 10.1088/0953-4075/36/5/304}
	{\bibfield  {journal} {\bibinfo  {journal} {J. Phys. B}\ }\textbf {\bibinfo
			{volume} {36}},\ \bibinfo {pages} {825} (\bibinfo {year} {2003})}\BibitemShut
	{NoStop}%
	\bibitem [{\citenamefont {Macr{\`{i}}}\ \emph {et~al.}(2013)\citenamefont
		{Macr{\`{i}}}, \citenamefont {Maucher}, \citenamefont {Cinti},\ and\
		\citenamefont {Pohl}}]{Macri2013}%
	\BibitemOpen
	\bibfield  {author} {\bibinfo {author} {\bibfnamefont {T.}~\bibnamefont
			{Macr{\`{i}}}}, \bibinfo {author} {\bibfnamefont {F.}~\bibnamefont
			{Maucher}}, \bibinfo {author} {\bibfnamefont {F.}~\bibnamefont {Cinti}}, \
		and\ \bibinfo {author} {\bibfnamefont {T.}~\bibnamefont {Pohl}},\ }\href
	{\doibase 10.1103/PhysRevA.87.061602} {\bibfield  {journal} {\bibinfo
			{journal} {Phys. Rev. A}\ }\textbf {\bibinfo {volume} {87}},\ \bibinfo
		{pages} {061602(R)} (\bibinfo {year} {2013})}\BibitemShut {NoStop}%
	\bibitem [{\citenamefont {Hsueh}\ \emph {et~al.}(2012)\citenamefont {Hsueh},
		\citenamefont {Lin}, \citenamefont {Horng},\ and\ \citenamefont
		{Wu}}]{Hsueh2012}%
	\BibitemOpen
	\bibfield  {author} {\bibinfo {author} {\bibfnamefont {C.-H.}\ \bibnamefont
			{Hsueh}}, \bibinfo {author} {\bibfnamefont {T.-C.}\ \bibnamefont {Lin}},
		\bibinfo {author} {\bibfnamefont {T.-L.}\ \bibnamefont {Horng}}, \ and\
		\bibinfo {author} {\bibfnamefont {W.~C.}\ \bibnamefont {Wu}},\ }\href
	{\doibase 10.1103/PhysRevA.86.013619} {\bibfield  {journal} {\bibinfo
			{journal} {Phys. Rev. A}\ }\textbf {\bibinfo {volume} {86}},\ \bibinfo
		{pages} {013619} (\bibinfo {year} {2012})}\BibitemShut {NoStop}%
	\bibitem [{\citenamefont {Roth}\ and\ \citenamefont
		{Burnett}(2003)}]{Roth2003}%
	\BibitemOpen
	\bibfield  {author} {\bibinfo {author} {\bibfnamefont {R.}~\bibnamefont
			{Roth}}\ and\ \bibinfo {author} {\bibfnamefont {K.}~\bibnamefont {Burnett}},\
	}\href {\doibase 10.1103/PhysRevA.67.031602} {\bibfield  {journal} {\bibinfo
			{journal} {Phys. Rev. A}\ }\textbf {\bibinfo {volume} {67}},\ \bibinfo
		{pages} {031602(R)} (\bibinfo {year} {2003})}\BibitemShut {NoStop}%
	\bibitem [{\citenamefont {Damski}\ \emph {et~al.}(2003)\citenamefont {Damski},
		\citenamefont {Zakrzewski}, \citenamefont {Santos}, \citenamefont {Zoller},\
		and\ \citenamefont {Lewenstein}}]{Damski2003}%
	\BibitemOpen
	\bibfield  {author} {\bibinfo {author} {\bibfnamefont {B.}~\bibnamefont
			{Damski}}, \bibinfo {author} {\bibfnamefont {J.}~\bibnamefont {Zakrzewski}},
		\bibinfo {author} {\bibfnamefont {L.}~\bibnamefont {Santos}}, \bibinfo
		{author} {\bibfnamefont {P.}~\bibnamefont {Zoller}}, \ and\ \bibinfo {author}
		{\bibfnamefont {M.}~\bibnamefont {Lewenstein}},\ }\href {\doibase
		10.1103/PhysRevLett.91.080403} {\bibfield  {journal} {\bibinfo  {journal}
			{Phys. Rev. Lett.}\ }\textbf {\bibinfo {volume} {91}},\ \bibinfo {pages}
		{080403} (\bibinfo {year} {2003})}\BibitemShut {NoStop}%
	\bibitem [{\citenamefont {Elstner}\ and\ \citenamefont
		{Monien}(1999)}]{Elstner1999}%
	\BibitemOpen
	\bibfield  {author} {\bibinfo {author} {\bibfnamefont {N.}~\bibnamefont
			{Elstner}}\ and\ \bibinfo {author} {\bibfnamefont {H.}~\bibnamefont
			{Monien}},\ }\href {\doibase 10.1103/PhysRevB.59.12184} {\bibfield  {journal}
		{\bibinfo  {journal} {Phys. Rev. B}\ }\textbf {\bibinfo {volume} {59}},\
		\bibinfo {pages} {12184} (\bibinfo {year} {1999})}\BibitemShut {NoStop}%
	\bibitem [{\citenamefont {Hohenadler}\ \emph {et~al.}(2011)\citenamefont
		{Hohenadler}, \citenamefont {Aichhorn}, \citenamefont {Schmidt},\ and\
		\citenamefont {Pollet}}]{Hohenadler2011}%
	\BibitemOpen
	\bibfield  {author} {\bibinfo {author} {\bibfnamefont {M.}~\bibnamefont
			{Hohenadler}}, \bibinfo {author} {\bibfnamefont {M.}~\bibnamefont
			{Aichhorn}}, \bibinfo {author} {\bibfnamefont {S.}~\bibnamefont {Schmidt}}, \
		and\ \bibinfo {author} {\bibfnamefont {L.}~\bibnamefont {Pollet}},\ }\href
	{\doibase 10.1103/PhysRevA.84.041608} {\bibfield  {journal} {\bibinfo
			{journal} {Phys. Rev. A}\ }\textbf {\bibinfo {volume} {84}},\ \bibinfo
		{pages} {041608(R)} (\bibinfo {year} {2011})}\BibitemShut {NoStop}%
	\bibitem [{\citenamefont {Heyl}\ \emph {et~al.}(2013)\citenamefont {Heyl},
		\citenamefont {Polkovnikov},\ and\ \citenamefont {Kehrein}}]{Heyl_2013}%
	\BibitemOpen
	\bibfield  {author} {\bibinfo {author} {\bibfnamefont {M.}~\bibnamefont
			{Heyl}}, \bibinfo {author} {\bibfnamefont {A.}~\bibnamefont {Polkovnikov}}, \
		and\ \bibinfo {author} {\bibfnamefont {S.}~\bibnamefont {Kehrein}},\ }\href
	{\doibase 10.1103/PhysRevLett.110.135704} {\bibfield  {journal} {\bibinfo
			{journal} {Phys. Rev. Lett.}\ }\textbf {\bibinfo {volume} {110}},\ \bibinfo
		{pages} {135704} (\bibinfo {year} {2013})}\BibitemShut {NoStop}%
	\bibitem [{\citenamefont {Titum}\ \emph {et~al.}(2019)\citenamefont {Titum},
		\citenamefont {Iosue}, \citenamefont {Garrison}, \citenamefont {Gorshkov},\
		and\ \citenamefont {Gong}}]{Paraj_2019}%
	\BibitemOpen
	\bibfield  {author} {\bibinfo {author} {\bibfnamefont {P.}~\bibnamefont
			{Titum}}, \bibinfo {author} {\bibfnamefont {J.~T.}\ \bibnamefont {Iosue}},
		\bibinfo {author} {\bibfnamefont {J.~R.}\ \bibnamefont {Garrison}}, \bibinfo
		{author} {\bibfnamefont {A.~V.}\ \bibnamefont {Gorshkov}}, \ and\ \bibinfo
		{author} {\bibfnamefont {Z.-X.}\ \bibnamefont {Gong}},\ }\href {\doibase
		10.1103/PhysRevLett.123.115701} {\bibfield  {journal} {\bibinfo  {journal}
			{Phys. Rev. Lett.}\ }\textbf {\bibinfo {volume} {123}},\ \bibinfo {pages}
		{115701} (\bibinfo {year} {2019})}\BibitemShut {NoStop}%
\end{thebibliography}


\end{document}